\documentclass[12pt,preprint]{aastex}

\newcommand{\ee}[1]{\mbox{${} \times 10^{#1}$}}
\newcommand{\hh}{\mbox{{\rm H}$_2$}}
\newcommand{\td}{\mbox{$T_d$} }
\newcommand{\msun}{\mbox{M$_\odot$}}
\newcommand{\Lsun}{\mbox{$L_{\odot}$}}

\shorttitle{Parameter Study of Dust and Gas Temperature}
\shortauthors{Urban et al.}                                                                                
\begin{document}

\title{A Parameter Study of the Dust and Gas Temperature in a Field of Young Stars}

\author{Andrea Urban\altaffilmark{1},
Neal J. Evans II\altaffilmark{1}, and Steven D. Doty \altaffilmark{2}}

\altaffiltext{1}{aurban@astro.as.utexas.edu, nje@astro.as.utexas.edu.\\Department of Astronomy, University of Texas, Austin, 
                 TX 78712, USA}

\altaffiltext{2}{doty@denison.edu.\\
Department of Physics and Astronomy, Denison University, Granville, OH 43023, USA}

\begin{abstract}
We model the thermal effect of young stars on their 
surrounding environment in order to understand clustered star formation.  
We take radiative heating of dust, dust-gas collisional heating, 
cosmic-ray heating, and molecular cooling into account.
Using Dusty, a spherical continuum radiative transfer code,
we model the dust temperature distribution around young stellar objects
with various luminosities and surrounding gas and dust density distributions. 
We have created a grid of dust temperature models, based on our modeling with
Dusty, which we can use to calculate 
the dust temperature in a field of stars with various parameters.
We then determine the gas temperature assuming energy balance. 
Our models can be used to make large-scale simulations
of clustered star formation more realistic. 

\end{abstract}


\section{Introduction}
Most of the stars in our galaxy form in groups or clusters 
\citep{lada}.
Therefore, in order to understand the star formation 
history, the shape of the mass function, 
and the formation of massive (M $\ga 5$ M$_{\odot}$) stars in our galaxy, 
the star formation process must be studied in its most 
common environment -- a cluster.  
As stars form from their initial reservoir of gas 
and dust, they  
interact with their environment and heat the 
surrounding material, thus affecting future star formation.  
One of the first effects a protostar has on its environment 
is radiative heating from the accretion luminosity and, 
subsequently, nuclear fusion.
The radiation efficiently heats the dust, which in turn 
heats the gas through collisions.  
Young stars also affect their environment via strong 
winds and ionization, 
but this only occurs when they are very massive and have evolved past
the very early stages of star formation.
We assume that the massive stars in our sample are very young and 
are accreting at very high rates 
($\dot{\rm{M}} \ga 1 \ee{-5} \msun / $yr).  
This high accretion rate allows
the infalling mass to absorb all of the stellar UV photons \citep{Churchwell}.

Many groups use large scale computer simulations to model clustered 
star formation.  
This is a complicated process requiring many assumptions in order to make the
problem tractable.
\cite{klessen} and  
Martel, Evans, \& Shapiro (2006) assume that the gas is isothermal. 
\citet{bbb} go beyond this assumption by using a barotropic equation of state. 
However, until recently, no one has included the effect of radiatively
heating the dust and gas by the stars formed in the simulation.  
\citet{krumholz} have included an approximate radiative 
transfer method, which works well in optically thick regions.  
Their method assumes that the gas temperature is 
equal to the dust temperature throughout their simulation.  
This approximation is only valid at high densities when the dust and gas are 
collisionally coupled.
Here we develop a method that explores the effect of radiative heating and
the dust and gas energetics for a range of optical depths and densities.

In our method we include various heating and cooling processes
to calculate the dust and gas temperature.  
Stars can heat dust grains more effectively than 
the gas because dust grains have broad-band absorption 
properties.  
Although we will not be explicitly modeling the motion 
or energy density of dust grains, we 
assume the dust and gas are well-mixed and the dust grains 
transfer energy to gas particles through collisions using
the energy transfer rate discussed in \citet{young}.  
The gas is heated by collisions with hot dust 
grains and cosmic rays.
It can cool through CO and other molecular line emission.   

In this paper, we calculate the dust and gas temperature
in a field of stars.  
The dust temperature around a single source is calculated
using a look-up table which we develop here.  
With this look-up table and an approximation
to the flux-temperature conversion, we calculate the dust
temperature in the field.
Our look-up table is needed since the calculation of a 
single dust temperature distribution can take longer than a 
minute on current desktops and would take a substantial 
fraction of a large-scale simulation's computations.  
Therefore we outline our method here which can be used to decrease
the time spent on the calculation of the dust temperature
in future studies of clustered star formation. 

With the calculated value of the dust temperature, we derive
the gas temperature field for a distribution of
stellar sources, as in a young stellar cluster.  
The effect that protostars have on heating their environment using a 
hydrodynamic and gravity simulation will be addressed in a future paper.
In this paper, we first discuss the calculation of the dust temperature 
for single and multiple sources (\S \ref{sec:analy}), 
then we describe our gas temperature calculation (\S \ref{sec:gas}), and 
finally, we show some dust and gas temperature distributions 
when multiple sources are present (\S \ref{sec:multiple}).

\section{Dust Temperature Calculation}\label{sec:analy}
We consider two methods of calculating the dust temperature when there are 
multiple heating sources. 
The first approach assumes radiative equilibrium after summing up the 
flux of multiple sources 
heating a dust grain. 
This approach makes simplifying assumptions about the dust absorption 
and emission properties. 
The second approach uses the one-dimensional spherical 
radiative transfer code Dusty (Nenkova, Ivezi{\'c}, \& Elitzur 2000),
to add up the energy density 
contributed by multiple sources; the dust temperature distribution is calculated separately 
for each source, then
the temperatures are converted to energy densities (assuming radiative equilibrium), 
which are summed together and then
converted back to a temperature.  
We first discuss the analytic 
approach (\S \ref{sec:1}), then the numerical 
approach to calculating the dust temperature (\S \ref{sec:2.2}),
and finally we compare and analyze the two approaches (\S \ref{sec:compare}). 

\subsection{Analytic Dust Temperature Calculation} \label{sec:1}
The rate of energy absorbed by a dust grain
in a field of $N$ stars is
\begin{equation}\label{eq:10}
\Bigg( \frac{dE}{dt}\Bigg)_{\rm{abs}} = \sum_{i=1}^{N} \frac{R_{*i}^2}{(\Delta \mathbf{r}_{*\it{i}})^2}
\int S_{\nu i} \pi a^2 Q_a(\nu) d\nu,
\end{equation}
where
$R_{*i}$ is the radius of star $i$, 
$S_{\nu i}$ is the flux density at the stellar surface of star $i$ (which we assume is a blackbody),
$Q_a$ is the dust grain's absorption efficiency
as a function of frequency, 
$\pi a^2$ is the projected 
surface area of the grain exposed to the star's light, 
and the separation between the star $i$ and the dust grain is
\begin{displaymath}
\Delta \mathbf{r}_{*\it{i}}=|\mathbf{r}_{*\it{i}}-\mathbf{r}|,
\end{displaymath}
where
$\mathbf{r}_{*\it{i}}$ is the location of star $i$ and 
$\mathbf{r}$ is the position of the dust grain.
Substituting in the Planck function for star $i$ at temperature $T_{*i}$ 
and assuming 
\begin{equation}\label{eq:1}
Q_a (\nu)= Q_a(\nu _o) \Bigg(\frac{\nu}{\nu_o}\Bigg)^{\beta _a},
\end{equation}
equation (\ref{eq:10}) becomes
\begin{equation}
\Bigg( \frac{dE}{dt}\Bigg)_{\rm{abs}} =\sum_{i=1}^{N} \frac{R_{*i}^2}{(\mathbf{r}_{*\it{i}})^2} \pi^2 a^2 
\frac{Q_a(\nu _o)}{\nu_o^{\beta _a}}
\frac{2h}{c^2} 
\Bigg(\frac{kT_{*i}}{h}\Bigg)^{4+\beta_a} 
I_{4+\beta_a}, 
\end{equation}
where
\begin{equation}
I_{4+\beta_a}=\Gamma(4+\beta_a) \zeta (4+\beta_a)
\end{equation}
and the functions $\Gamma(x)$ and $\zeta(x)$ are defined as the gamma and Riemann zeta
function.


The same assumptions can be made about the emission of the grain 
except that the grain emits in all directions 
(i.e. $\pi a^2$ becomes $4 \pi a^2$), 
the grain is emitting instead of the star
(i.e. $T_{*i}$ becomes $\td$), 
and the dust grain's emission efficiency is
\begin{equation}\label{eq:be}
Q_e (\nu)= Q_e(\nu _o) \Bigg(\frac{\nu}{\nu_o}\Bigg)^{\beta _e}.
\end{equation}
These assumptions give
\begin{equation}
\Bigg( \frac{dE}{dt}\Bigg)_{\rm{emis}} = \sum_{i=1}^{N}
 4 \pi^2 a^2 
\frac{Q_e(\nu _o)}{\nu_o^{\beta _e}}
\frac{2h}{c^2} 
\Bigg(\frac{kT_{d}}{h}\Bigg)^{4+\beta_e} 
I_{4+\beta_e}.
\end{equation}
Assuming radiative equilibrium and 
removing the dependence on stellar radius with $L_*=4\pi R_*^2 \sigma T_*^4$, 
$\td$ at position $\mathbf{r}$ is given by
\begin{equation}\label{eq:td}
\td(\mathbf{r}) = \Bigg[ \frac{1}{4} \frac{Q_a(\nu_o)}{Q_e(\nu_o)} \frac{\nu_o^{\beta_e}}{\nu_o^{\beta_a}}
\frac{I_{4+\beta_a}}{I_{4+\beta_e}}
\Big( \frac{k}{h} \Big)^{\beta_a - \beta_e}
\frac{1}{4\pi\sigma}
\sum_{i=1}^{N}
\frac{L_{*i}T_{*i}^{\beta_a}}{(\Delta \mathbf{r}_{*\it{i}})^2}
\Bigg]^{1/(4+\beta_e)}
\end{equation}

If we assume that grains absorb like graybodies ($\beta_a=0$)
where stars emit most of their luminosity (UV), then
$Q_a (\nu)= Q_a(UV)$.
The value of $Q_e$ is given at $125 \mu$m with respect to $Q_a$.  
Using this approximation, we obtain
\begin{equation}\label{eq:td}
\td(\mathbf{r}) = \Bigg[ 3.89\ee{10} \frac{Q_a(UV)}{Q_e(125\mu m)} \frac{115^{\beta_e}}{I_{4+\beta_e}}
\sum_{i=1}^{N}
\frac{L_{*i}/1L_{\bigodot}}{(\Delta  \mathbf{r}_{*\it{i}}/1AU)^2}
\Bigg]^{1/(4+\beta_e)}
\end{equation}
Notice that the dependence on stellar temperature disappears in this case.

The value of $Q_a(UV) / Q_e(125\mu m)$, 
where the UV wavelength range is $0.15 \mu m - 0.30 \mu m$, 
can be calculated from observations or from a dust grain model.  
We compare various values from the literature to the value 
derived from the dust we use in our models (see Table \ref{tab:a}).
Our dust model is a combination of OH5 dust  \citep{oh5} 
and \citet{pollack} dust as described in \citet{chad}.
For OH5 dust, we calculate the value of $Q_a(UV) / Q_e(125\mu m)$ 
assuming a $10,000$K blackbody.  
$Q_{a}$ is the stellar flux weighted average absorption efficiency
of the dust and $\sigma _{\rm{abs}}(\lambda)$ is the absorption cross-section for a dust grain.  
Therefore,
\begin{equation}
\frac{Q_a(UV) }{Q_e(125 \mu m)} = \Bigg( \frac{
\sum^{0.3 \mu m}_{\lambda = 0.15 \mu m} F_{\lambda} \sigma_{\rm{abs}}(\lambda) }
{\sum^{0.3 \mu m}_{\lambda = 0.15 \mu m} F_{\lambda} } \Bigg)
\Bigg/
{\sigma_{\rm{abs}}(125 \mu m)} = 253.
\end{equation}

\begin{figure}[!ht]
\epsscale {1.0}
\plotone{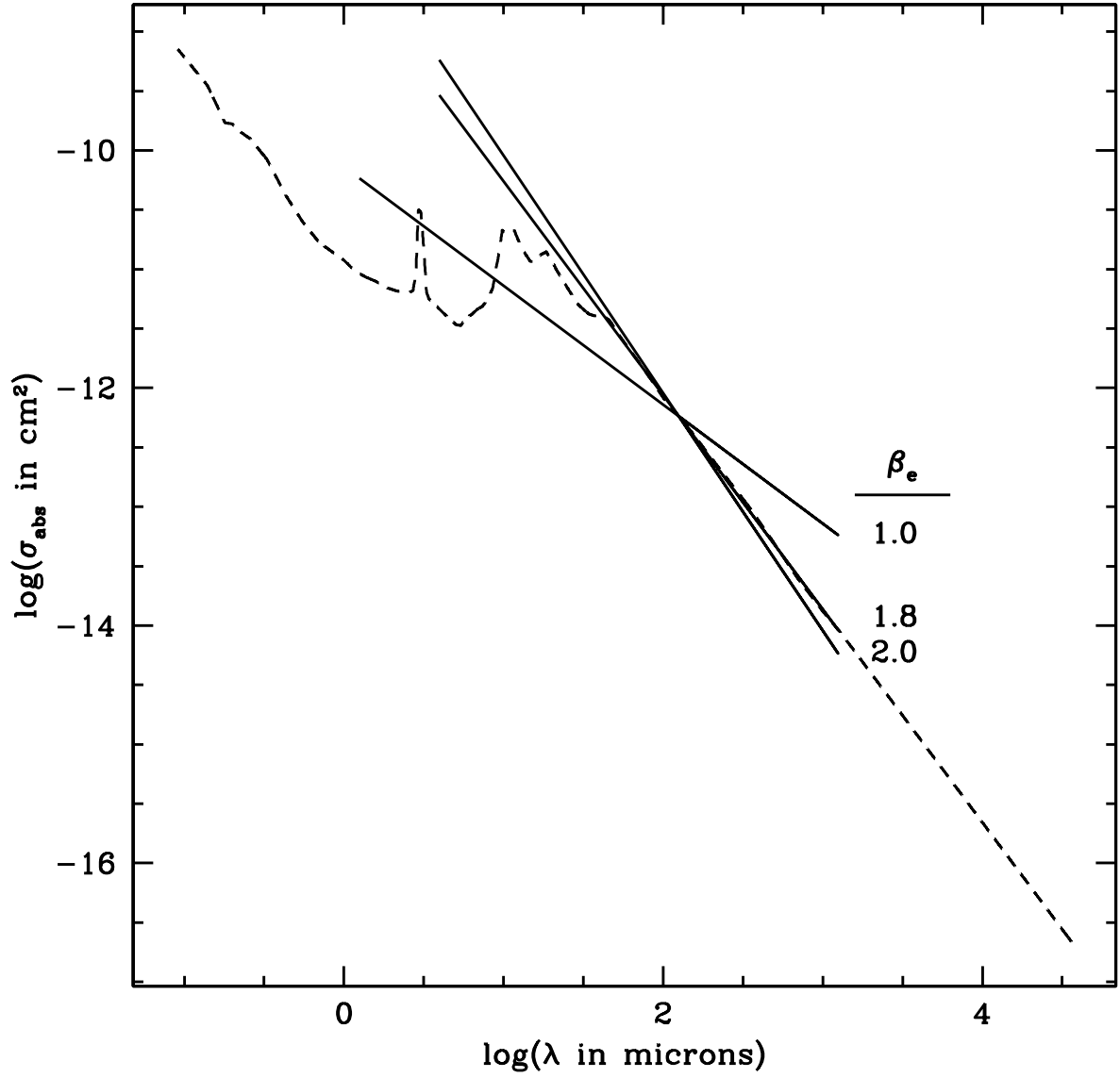}
\caption{OH5 dust properties. Dashed line shows the variation of 
cross-section with wavelength for OH5 dust.  
The solid lines show the different values of 
$\beta_e$ normalized at $125 \mu $m that we consider 
in Table \ref{tab:a}.
\label{fig:sig}}
\end{figure}

Now we consider the various dust models in Table \ref{tab:a}
and fix the value of $\beta_e$, the dust grain's emission efficiency exponent. 
From equation (\ref{eq:td}), the form of the dust temperature profile becomes
\begin{equation}\label{eq:basic}
\td = K(\beta_e)
\Bigg[ \sum_{i=1}^{N}
\frac{L_{*i}/1L_{\odot}}{(\Delta \mathbf{r}_{*\it{i}}/1 AU)^2}
\Bigg]^{1 / (4+\beta_{e})},
\end{equation}
where $\beta_{e}$
is fixed and the values of $K(\beta_e)$ for each model
are given in Table \ref{tab:a}.  
In Figure \ref{fig:sig} we compare OH5 dust to an analytic approximation
(see equation \ref{eq:be}) normalized at $125 \mu $m and vary the value 
of $\beta_e$. 
The line with $\beta_e=1.8$ fits well at 
long wavelengths but not at shorter wavelengths.
The opposite is true for $\beta_e=1.0$.  
Since we expect long wavelength emission to dominate these clouds, we adopt an 
analytic approximation of OH5 dust with $\beta_e=1.8 $ 
as our ``Analytic Solution'' in the following sections. 

\subsection{Numerical Dust Temperature Calculation}\label{sec:2.2}
Another method of calculating the dust temperature around multiple sources uses
the code Dusty which we have set up with OH5 dust opacities \citep{oh5} using the 
method described in \citet{chad}.
Dusty is a one-dimensional spherical radiative transfer code \citep{dusty}.
Once a dust temperature distribution is derived around a single source,
we use it to estimate the dust temperature around multiple 
sources, using some of the assumptions in the previous method,
which we explain in more detail below.  
Using Dusty to calculate the dust temperature around a young
star is the most accurate solution; 
however, this program can take over 1 minute
to run for low optical depths with yet longer run times for larger $\tau$.
Therefore, we calculate the dust temperature profile for various
combinations of luminosity, outer radius, and density profile to 
create a look-up table.  
(The parameters we consider are discussed in \S \ref{sec:par}.)
We then assume that the dust temperature profile around a single 
source is of the form given in equation (\ref{eq:td}), i.e.
\begin{equation}\label{eq:A}
\td (\Delta \mathbf{r}_{*\it{i}})= K_i \Bigg( \frac{L_{*i}}
{(\Delta \mathbf{r}_{*\it{i}})^2}\Bigg)^{1/(4+\beta_i)},
\end{equation}
where $K$ and $\beta$ are functions of the density profile and dust properties.

This assumption is valid when the dust is optically thin.  Although
the gas and dust are denser closer to the central source and likely to be optically thick, 
we are mainly interested in the temperature distribution far from the central
source where the physical processes we consider are dominant 
and the dust and gas are optically thin to the cloud exterior.
For example, close to a forming star [$r \la 1000$ AU \citep{mcC}]
there may be a disk, which is 
not represented by our assumption of spherical symmetry.  Therefore, it is not
useful to model the dust temperature close to the star because it will be
affected by the presence of a disk.

For each set of parameters, 
we run Dusty and solve for a value of $K$ and $\beta$ 
(see \S \ref{sec:par}).
Since we are only interested in the dust temperature far from the 
source, we fit the outer 25\% of the dust temperature profile 
in $\log T- \log r$ space using least-squares fitting 
in order to determine the values of $\beta$ and $K$ in equation (\ref{eq:A}).
We call these values of $\beta$ and K our ``Fit Solution''.
In order to calculate the dust temperature of a region heated by more than one protostar, 
we add up the flux at the region of interest using
\begin{equation}\label{eq:B}
F(\mathbf{r})=\sum^N_{i=1} \frac{L_{*i}}{4 \pi (\Delta \mathbf{r}_{*\it{i}})^2}
\end{equation}
Combining equations (\ref{eq:A}) and (\ref{eq:B}), we derive 
\begin{equation}\label{eq:Q}
F(\mathbf{r})=\sum^N_{i=1} \frac{(\td(\Delta \mathbf{r}_{*\it{i}})/K_i)^{4+\beta_i}}{4 \pi }. 
\end{equation}
In general, we assume 
\begin{equation}\label{eq:R}
F(\mathbf{r})=\frac{(\td(\mathbf{r})/\bar{K})^{4+\bar{\beta}}}{4 \pi}
\end{equation}
where $\bar{K}$ and $\bar{\beta}$ are defined as the flux-weighted 
averages of the $K$ and $\beta$ values that 
contribute to the flux at point $\mathbf{r}$, i.e. 
\begin{equation}
\bar{\beta} = \frac{\sum \beta_i L_{*i} / 4 \pi \Delta \mathbf{r}_{*\it{i}}^2}
{\sum L_{*i} / 4 \pi \Delta \mathbf{r}_{*\it{i}}^2}
\end{equation}
and
\begin{equation}
\bar{K} = \frac{\sum  K_iL_{*i} / 4 \pi \Delta \mathbf{r}_{*\it{i}}^2}
{\sum L_{*i} / 4 \pi \Delta \mathbf{r}_{*\it{i}}^2},
\end{equation}
where the sums are from $i=1$ to the total number of stars, $N$. 

Therefore, equating (\ref{eq:Q}) and (\ref{eq:R}) 
and solving for $\td(\mathbf{r}$) gives, 
\begin{equation}\label{eq:tdsum}
\td(\mathbf{r}) = \bar{K}\Bigg[
\sum^N_{i=1} \Bigg(\frac{T(\Delta \mathbf{r}_{*\it{i}})}
{K_i}\Bigg)^{4+\beta_i}\Bigg]^{1/ (4+ \bar{\beta})}.
\end{equation}
Using equation (\ref{eq:A}), we obtain
\begin{equation}\label{eq:tdsum1}
\td(\mathbf{r}) = \bar{K}\Bigg[
\sum^N_{i=1} \frac{L_{*i}}{\Delta \mathbf{r}_{*\it{i}}^2},
\Bigg]^{1/ (4+ \bar{\beta})}  
\end{equation}
the equation we used to calculate the dust temperature in a field of sources.

%

\subsubsection{Parameter Space}\label{sec:par}
We assume the dust and gas are well-mixed and have the same 
density profile, offset only by the dust to gas mass ratio ($\eta_{dg}$) 
described in \S \ref{sec:gas}, i.e. 
$\rho_{\rm{dust}}=\eta_{dg} \rho_{\rm{gas}}$
We assume $N_{\rm{H}_2}/N_{\rm{He}}=5$, which gives $\mu=2.33$.
The density profile of the gas is parameterized with $n_o$ and $\alpha$ using
\begin{equation}
n_{gas}=n_o \bigg( \frac{r}{1000AU} \bigg)^{-\alpha}.
\end{equation}

We choose an inner radius of the dust distribution to be fixed at $30$AU.  
This number is smaller than the average sizes of more evolved disks 
[i.e. \citet{mcC} find disk diameters of 50-1000 AU for disks in Orion], 
which would be considered sub-grid physics in a simulation of 
clustered star formation with moderate resolution.
However, we find that the choice of the inner radius does not greatly
affect the values of $\beta$ and $K$ for our models since we are not modeling
the increase in temperature where the dust becomes optically thick.  

The luminosity required to reach a dust destruction temperature of $1500$ K at 
$30AU$ using the Analytic Solution is $9.58\ee{6} \Lsun$.  
This is outside our sample range of luminosities; 
therefore we can neglect the effects of dust destruction.
We also vary the outer radius, $r_{\rm{out}}$, to obtain a range of values.  
We find that our values of $K$ and $\beta$ do not strongly depend on $r_{\rm{out}}$.

In all cases, the stellar input spectrum is assumed to be a 
blackbody with temperature $T_{*} = 10,000K$.  
This value does not strongly influence the output temperature
distribution at outer radii since the light is
quickly reprocessed by the dust to longer wavelengths.

\begin{figure}[!ht]
\epsscale{1.2}
\plottwo{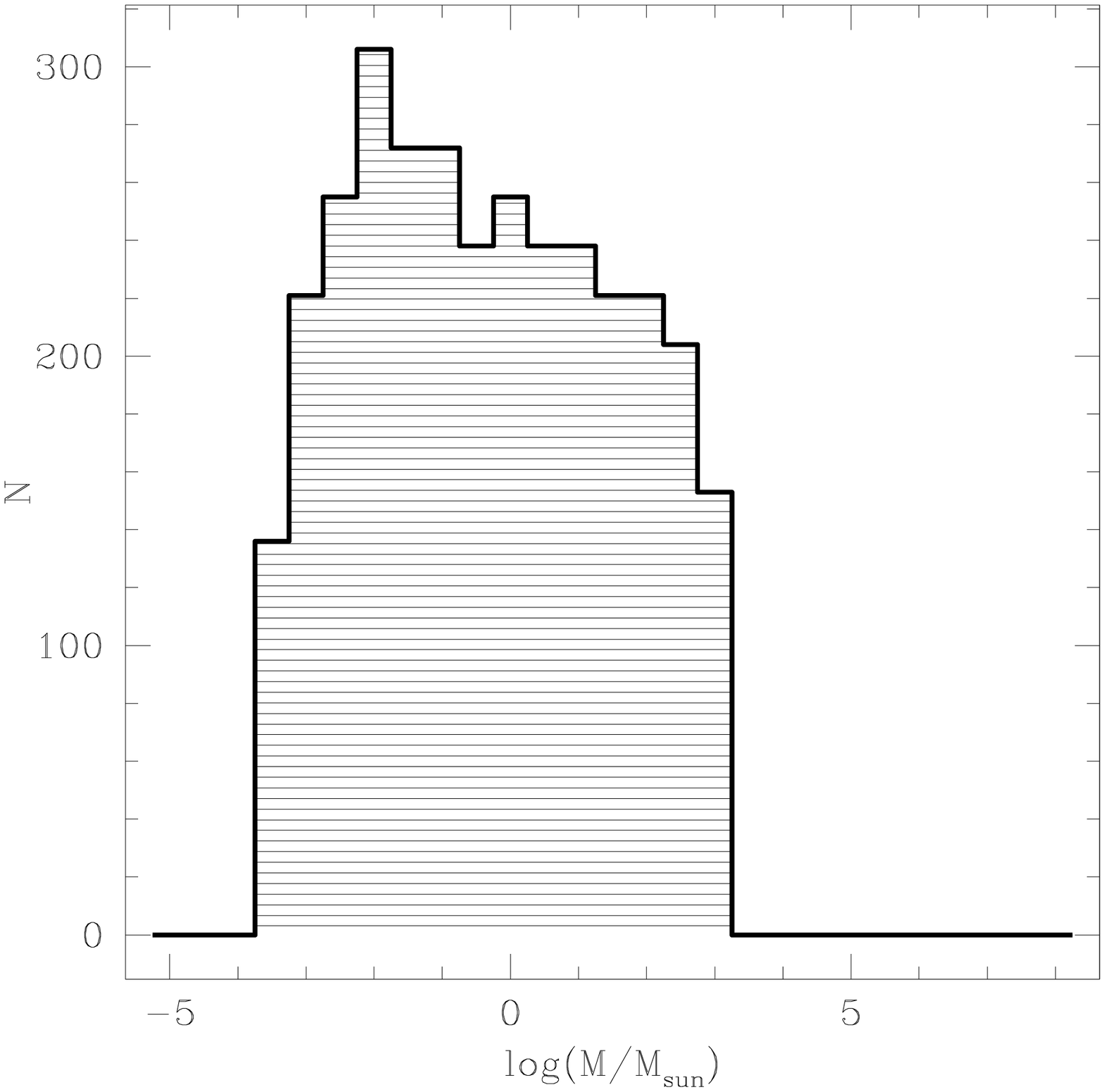}{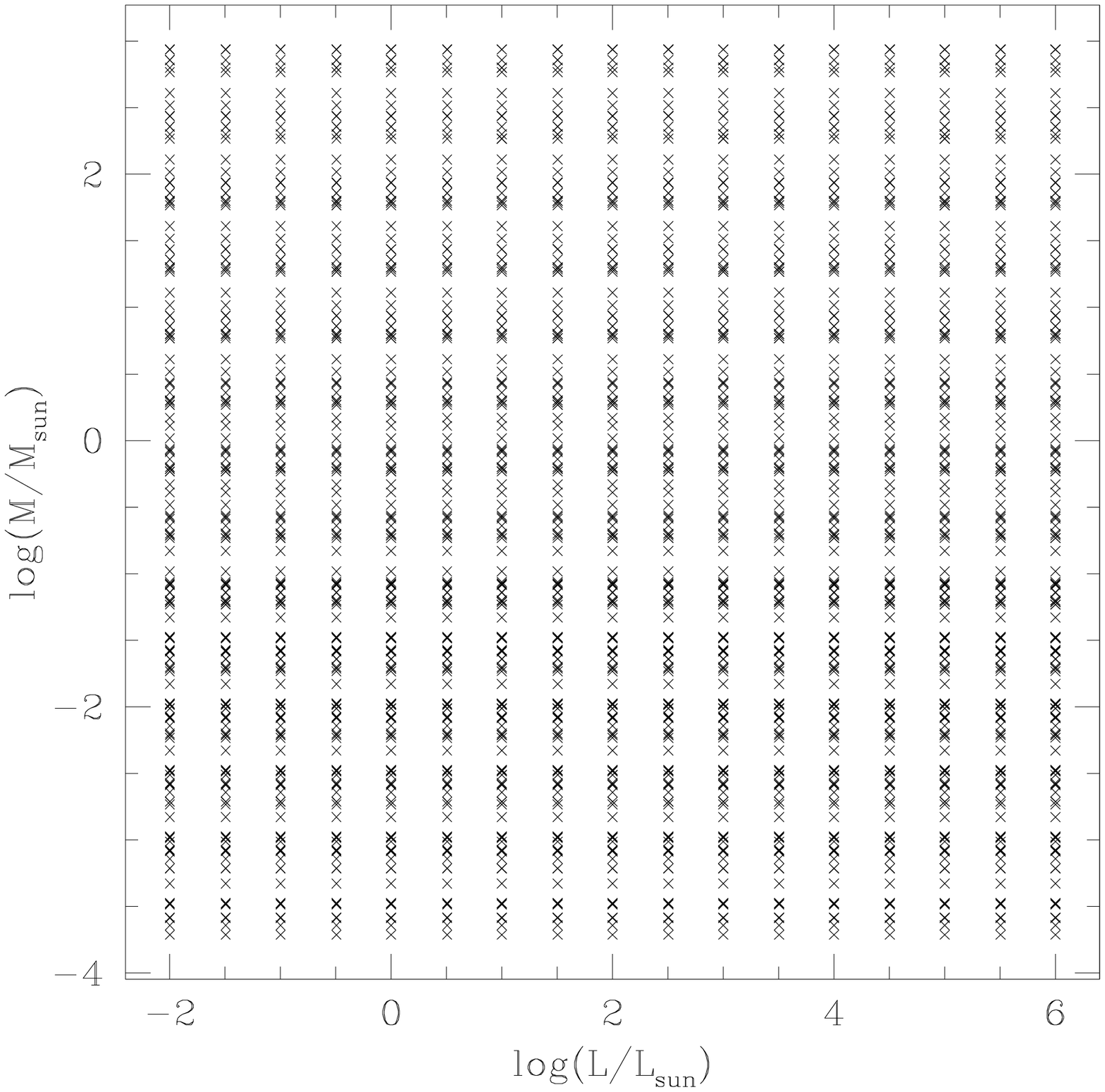}
\caption{Masses and luminosities of models in our parameter space.  
Figure on the left shows a histogram of the masses of the clouds surrounding
a source. On the right, the relationship between the mass and luminosity of 
all our models is plotted.  
For every density distribution in our sample, we have models which correspond
to all of the luminosity values in our parameter space.
\label{fig:mass}}
\end{figure}

\begin{figure}[!ht]
\epsscale{1.2}
\plottwo{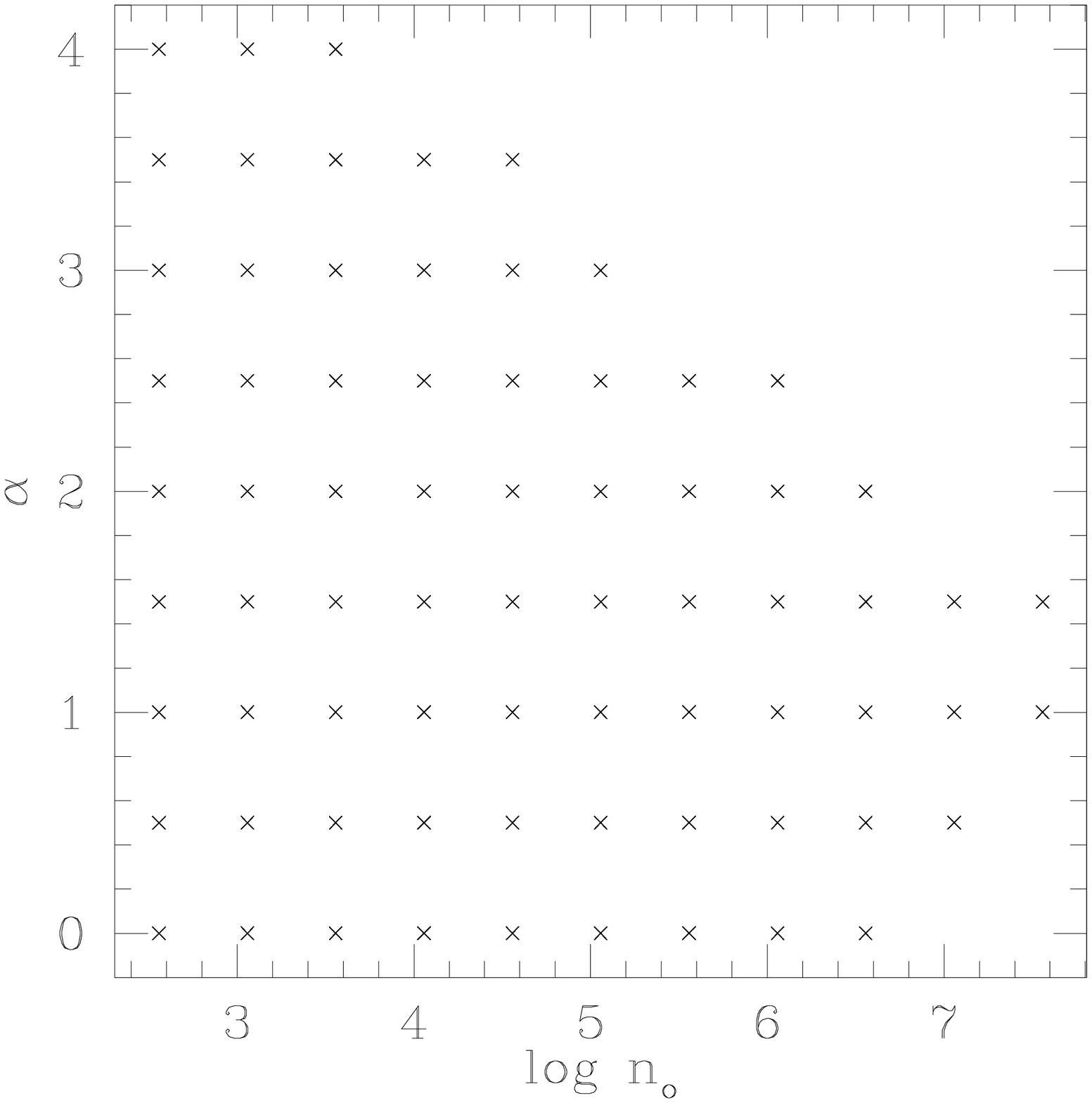}{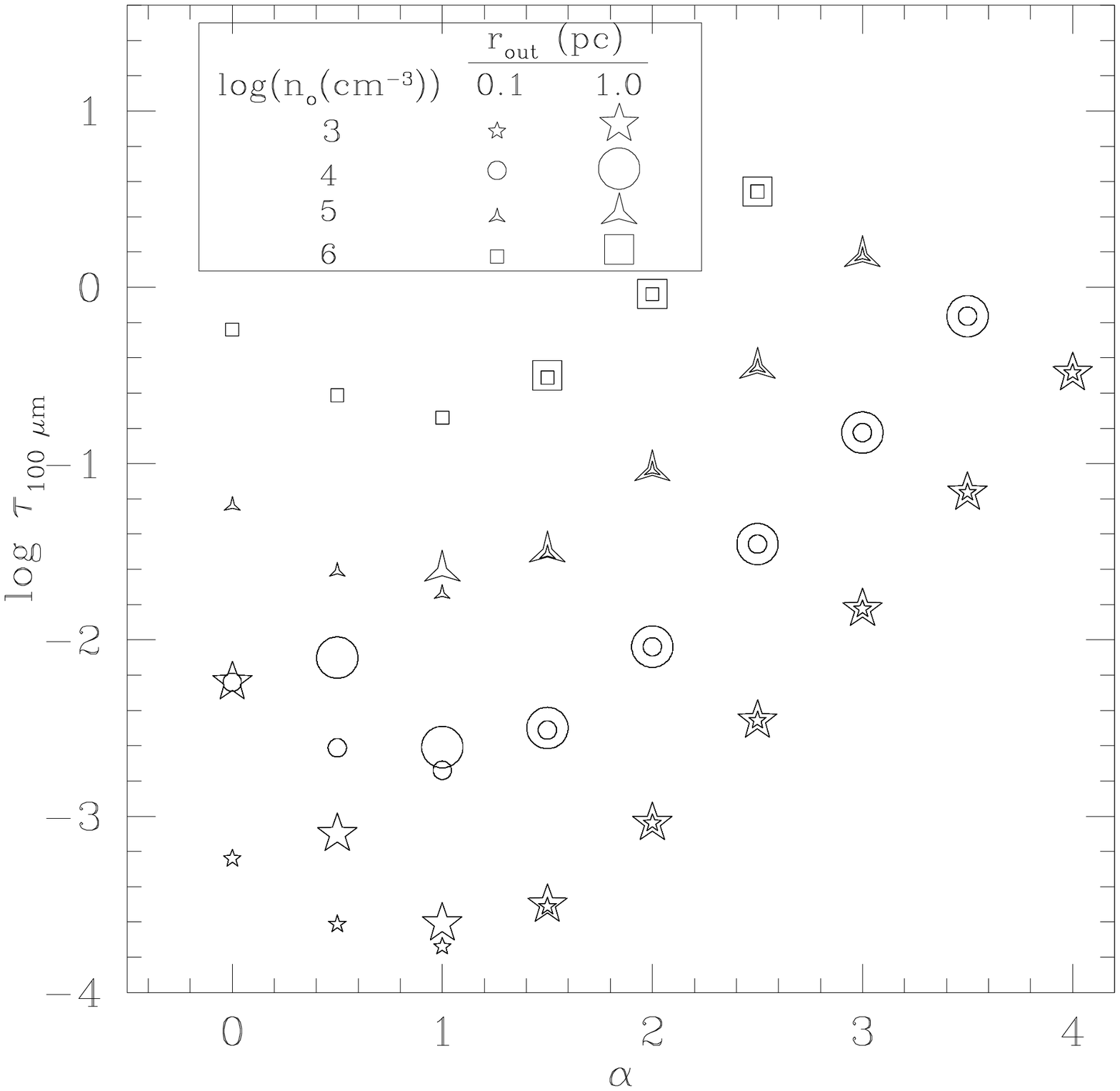}
\caption{Relationship between $\alpha$, $\tau$, and $n_o$.
The figure on the left shows the relationship between $\alpha$ and $n_o$
for all our models.  Some models at low $\alpha$ with high values of 
$n_o$ are not in our sample due to the maximum mass criterion.  
Models missing in the top right corner of the figure at high $n_o$ and high
$\alpha$ are missing due to the criterion which sets the maximum density at
the inner radius.
On the right, we show how $\tau$ varies with 
$\alpha$, $n_o$, and $r_{out}$ for our models.  
At a fixed value of $n_o$, as $\alpha$ increases (and $\alpha > 1$), 
$\tau$ increases as well due to the sharp density increase in the profile.
For lower values of $\alpha$ ($\alpha < 1$), $\tau$ begins to increase
again, but this increase depends on amount of material included in the 
profile at the edge, i.e. the value of $r_{out}$.  
\label{fig:alpha}}
\end{figure}

We model the entire parameter space listed in Table \ref{tab:2par}
with two exceptions.
First, we limit the density at the inner radius to be less than 
$10^{10}$cm$^{-3}$, because at 
higher densities dynamical effects may become important since the 
free-fall time becomes small as the density increases. 
Second, we limit the mass of the envelope to be less than $\sim 1000$ $\msun$ 
since a larger envelope would likely produce a cluster of stars 
(assuming a star formation efficiency of 10\% and a maximum stellar 
mass of 100 $\msun$)
which would break the assumption of spherical symmetry.
Therefore, combinations of large $\alpha$ and 
large $n_o$ may not be represented in our models.  
Based on these restrictions, of the 5049 possible models in our 
parameter space, we model 3231 or 64\% of them.
In Figure \ref{fig:mass} we show the luminosities
and masses modeled in our parameter space.  
Figure \ref{fig:alpha} shows the relationship 
between the values of $\tau$, $\alpha$ and $n_o$ for the models 
in our parameter space.

\subsection{Comparison of Dust Temperature Calculation Methods}\label{sec:compare}
\begin{figure}[!ht]
\epsscale{1.2}
\plottwo{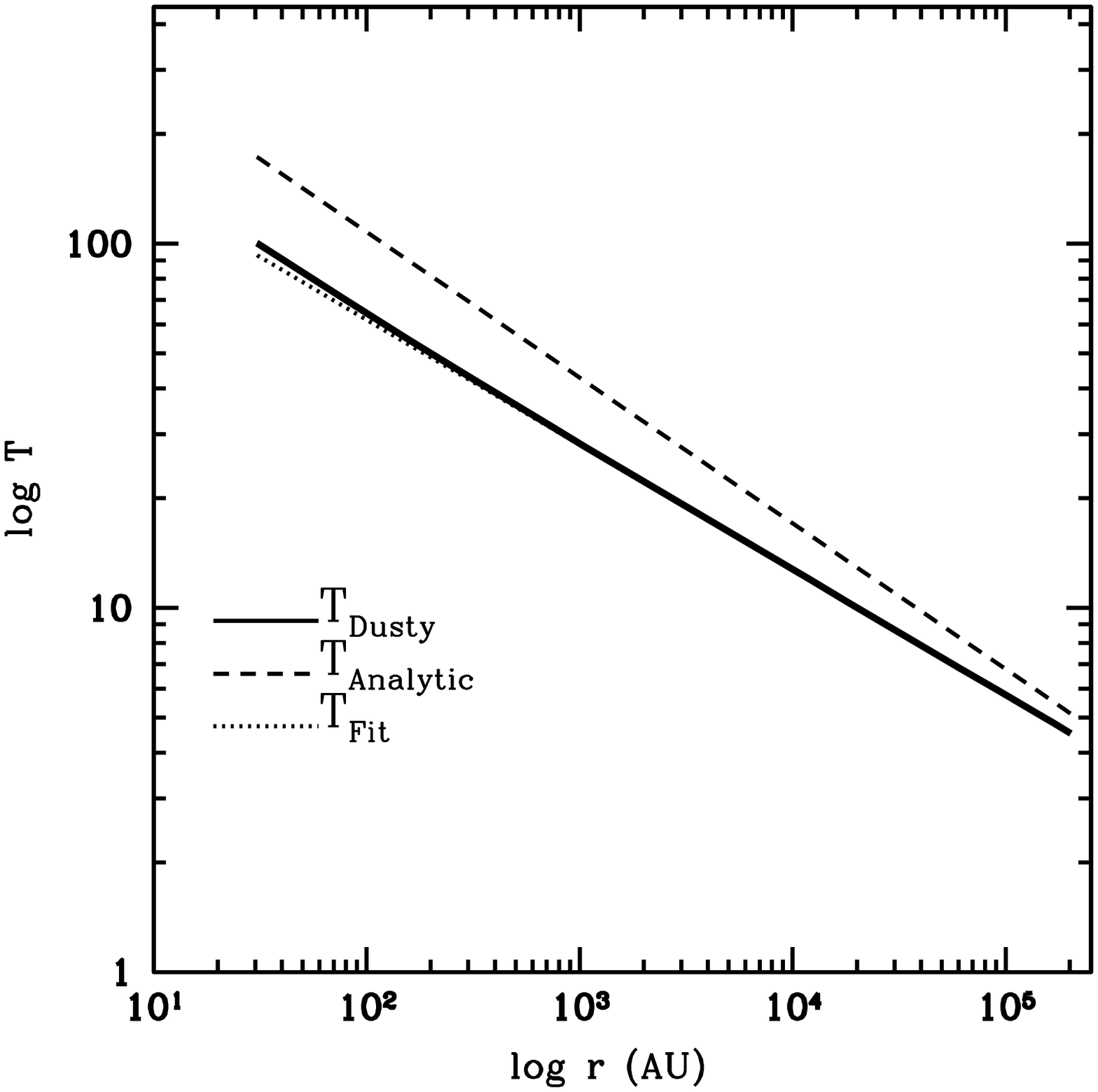}{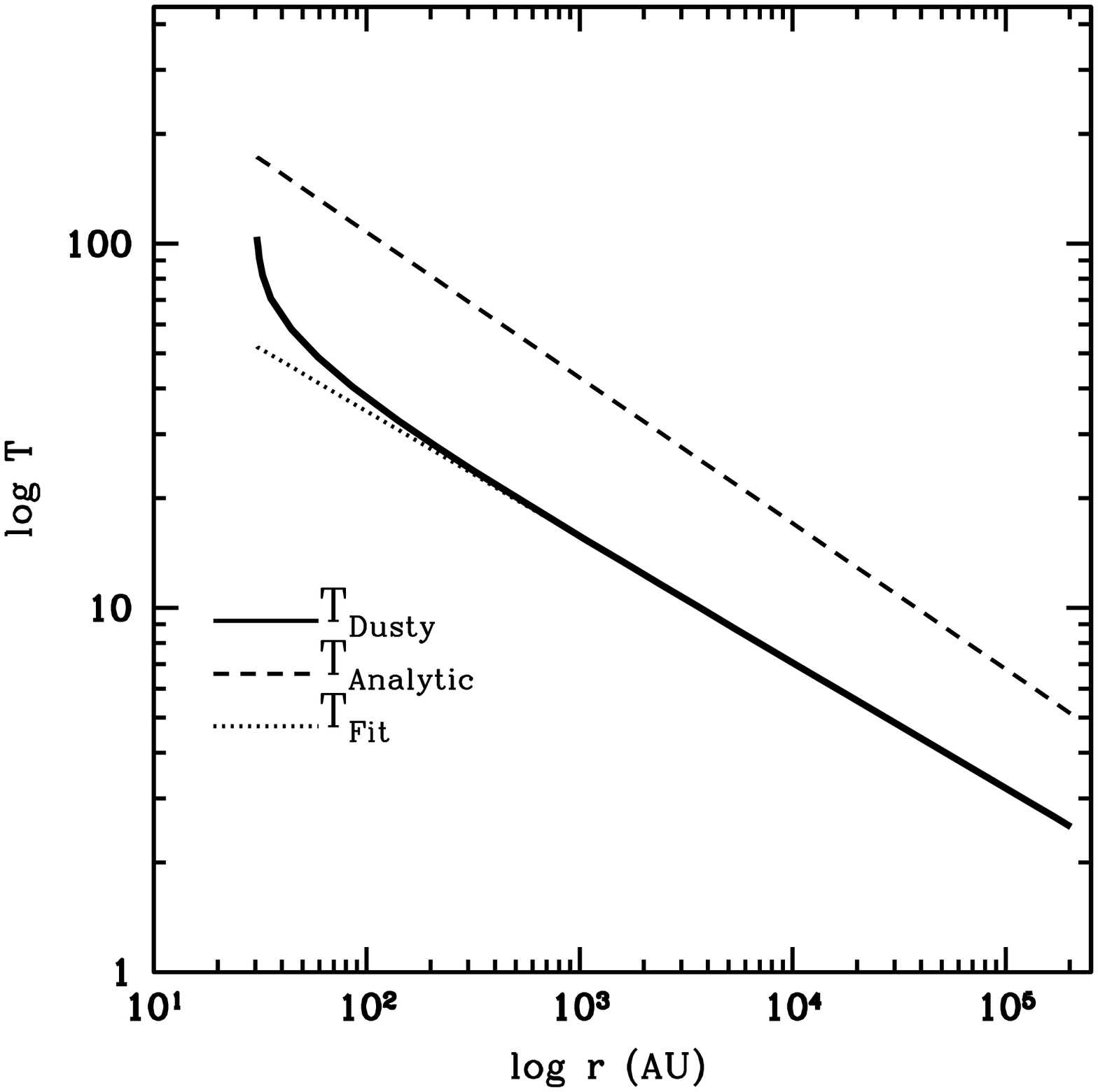}
\caption{Comparison between temperature distributions for two different models with a low and high fiducial density.
For both models, $L=1\Lsun$, $\alpha = 2$, and $r_{out} = 1$ pc.  
The figure on the left has log $n_o=2.5$ and $\tau = 2.894\ee{-4}$. 
The fit parameters are $K=300.79$ and $\beta =1.822$. 
The figure on the right has log $n_o=5.5$ and $\tau = 2.894\ee{-1}$ with 
$K=169.5$ and $\beta =1.799$ .
Although both of these models are optically thin, as assumed in the 
Analytic Solution, it is clear that it is not a good description of 
the dust temperature.  If we were to change the value of $K$ assumed
in the Analytic Solution, we might be able to fit the exact 
solution provided by Dusty for one model, but it would then not fit 
for another model.  
\label{fig:fidno}}
\end{figure}

\begin{figure}[!ht]
\epsscale{1.2}
\plottwo{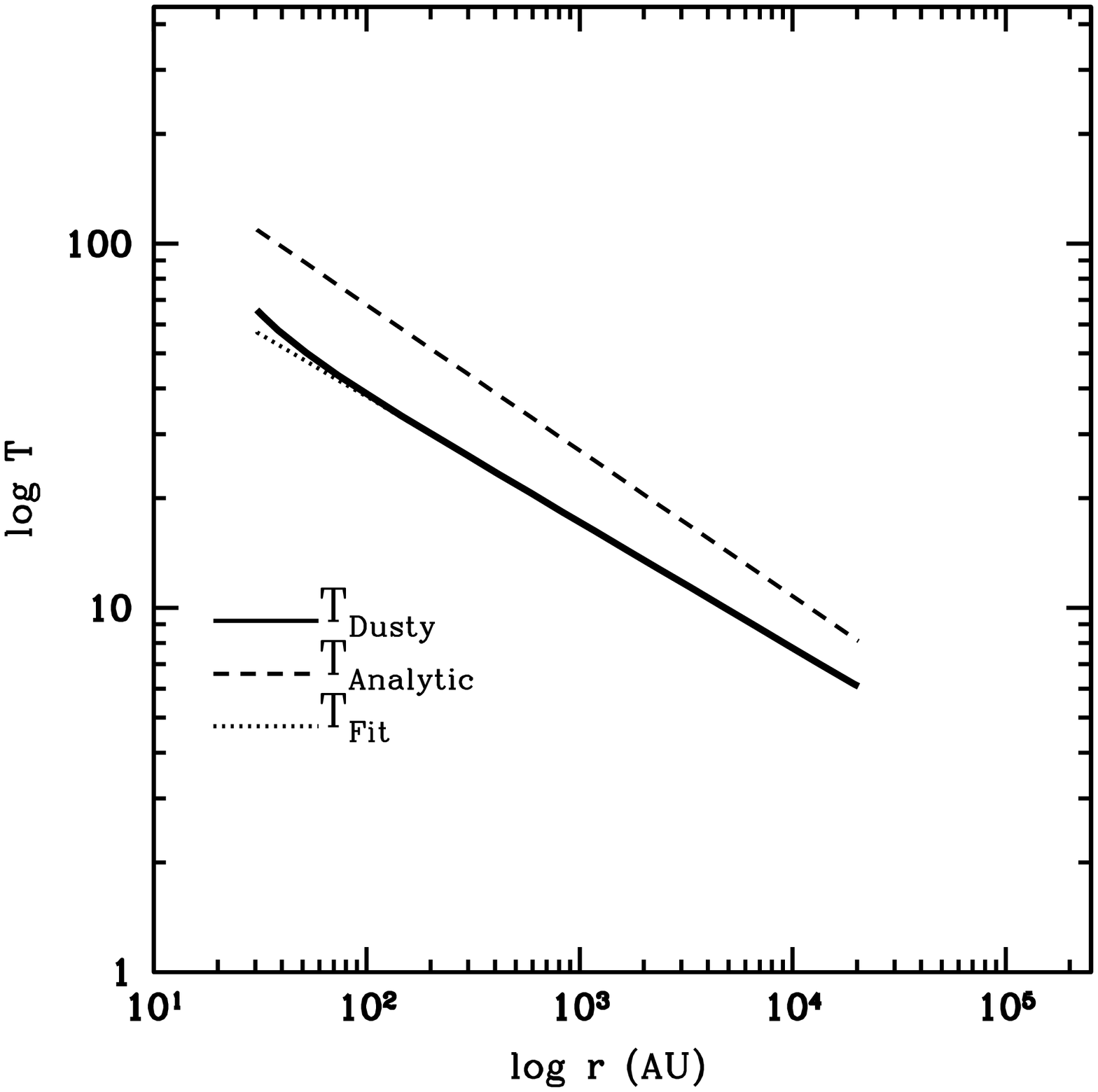}{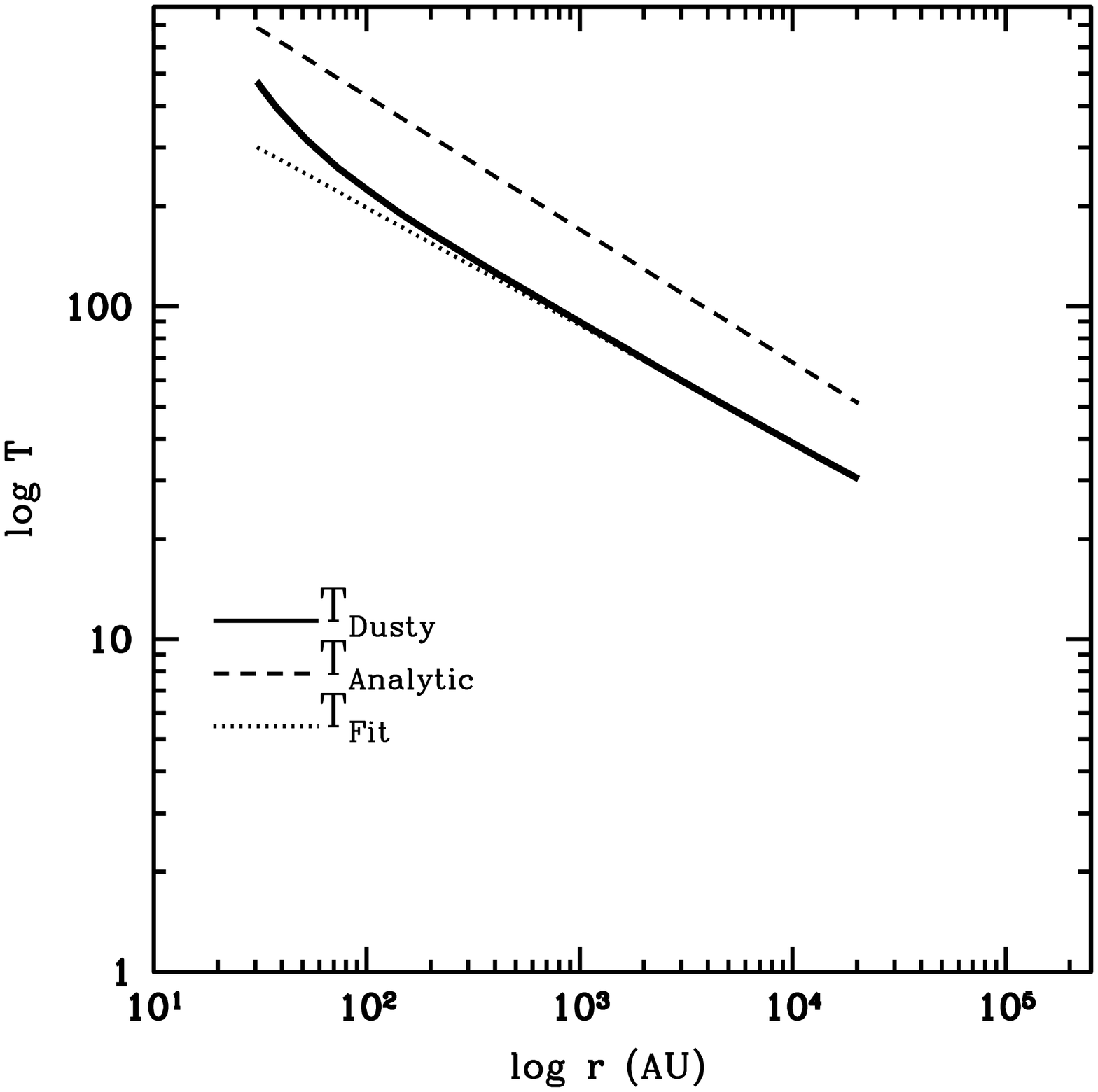}
\caption{Comparison between temperature distributions for two
different models with a low and high luminosity.
For both models, $\alpha = 3$, and $r_{out} = 0.1$ pc, log $n_o=2.5$.  
The figure on the left has log $L/\Lsun=-1$.
The fit parameters are $K=276.23$ and $\beta =1.802$ 
The figure on the right has log $L/\Lsun=3$ with
$K=297.5$ and $\beta =1.658$.
These models show how increasing the luminosity
increases the overall dust temperature as well as making the 
optically thick region near the center extend farther out.
\label{fig:fidL}}
\end{figure}

Figures \ref{fig:fidno} and \ref{fig:fidL} compare the two methods of 
calculating the dust temperature around a single source.
The Analytic Solution uses OH5 dust parameters with $\beta _e=1.8$ as described in 
\S \ref{sec:1} which best matches the dust properties used in the 
Dusty Solution in Figure \ref{fig:sig}.

Figure \ref{fig:fidno} shows that the Analytic Solution captures the \emph{shape} of the 
Dusty temperature profile, but at high $\tau$ it overestimates the \emph{magnitude}.
If we only vary the luminosity (see Figure \ref{fig:fidL})
the offset of the Analytic Solution to the Dusty Solution changes, 
therefore a simple adjustment to the Analytic Solution based on 
luminosity would not correct the Analytic Solution.
The Fit Solution described in \S \ref{sec:2.2} appears to be the best fit, 
as shown in Figures \ref{fig:fidno} and \ref{fig:fidL}.

\begin{figure}[!ht]
\epsscale {1.2}
\plottwo{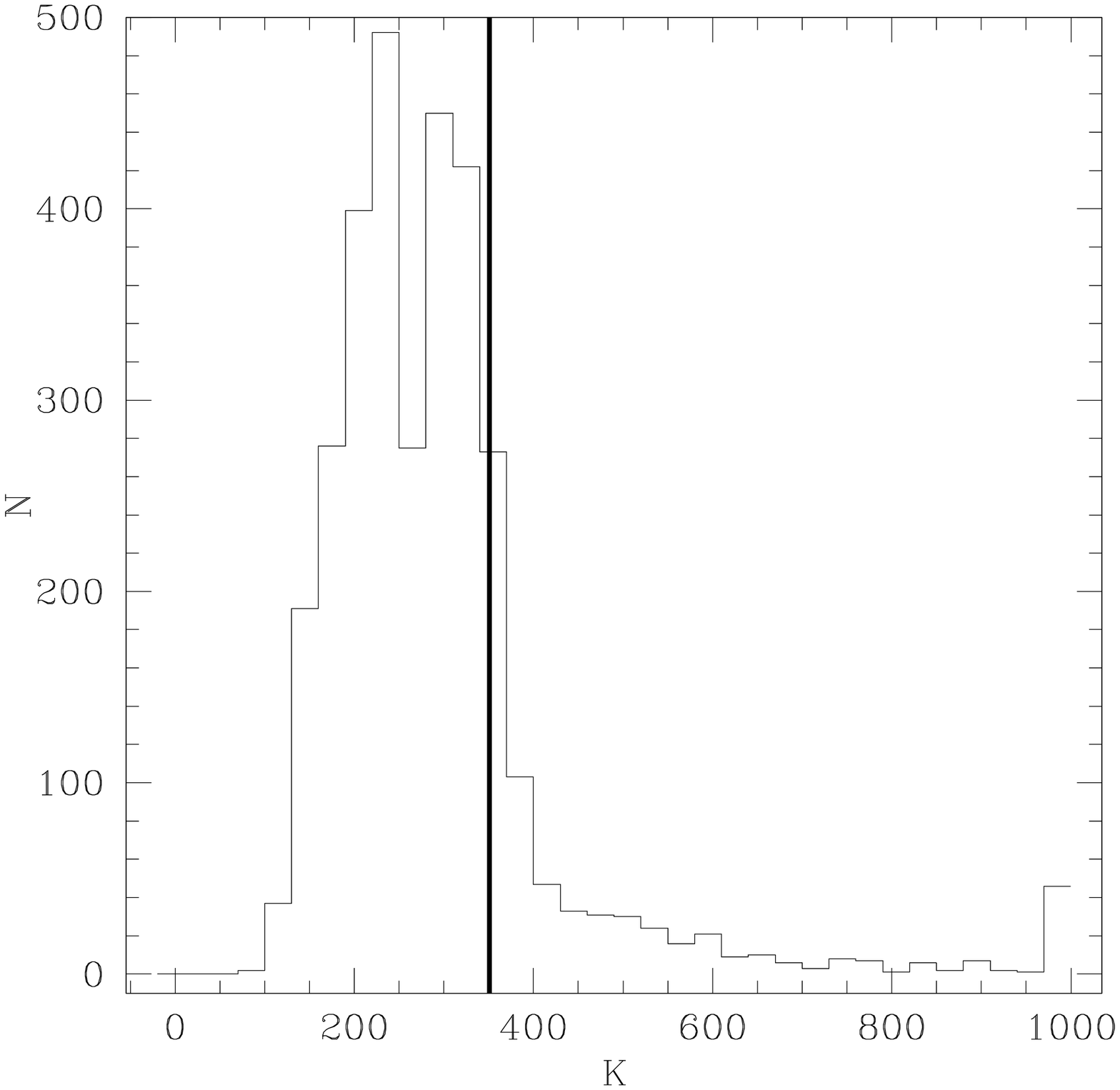}{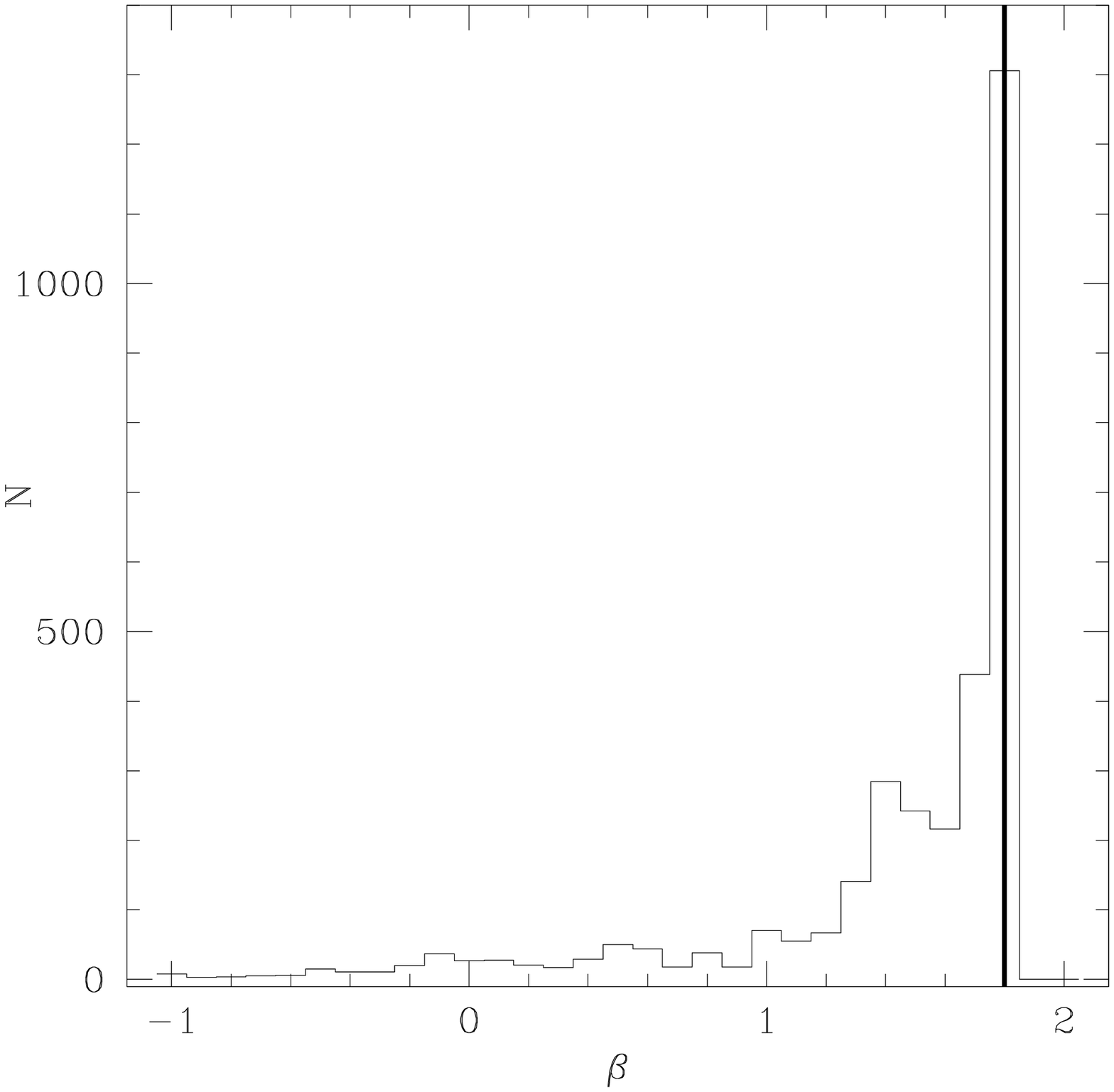}
\caption{Figures shows the range of K and $\beta$ for the chosen
parameters in Table \ref{tab:2par}.  The vertical line marks the values of $K$ and
$\beta_e$ in the Analytic Solution. 
\label{fig:Kbeta}}
\end{figure}

\begin{figure}[!ht]
\epsscale {1.2}
\plottwo{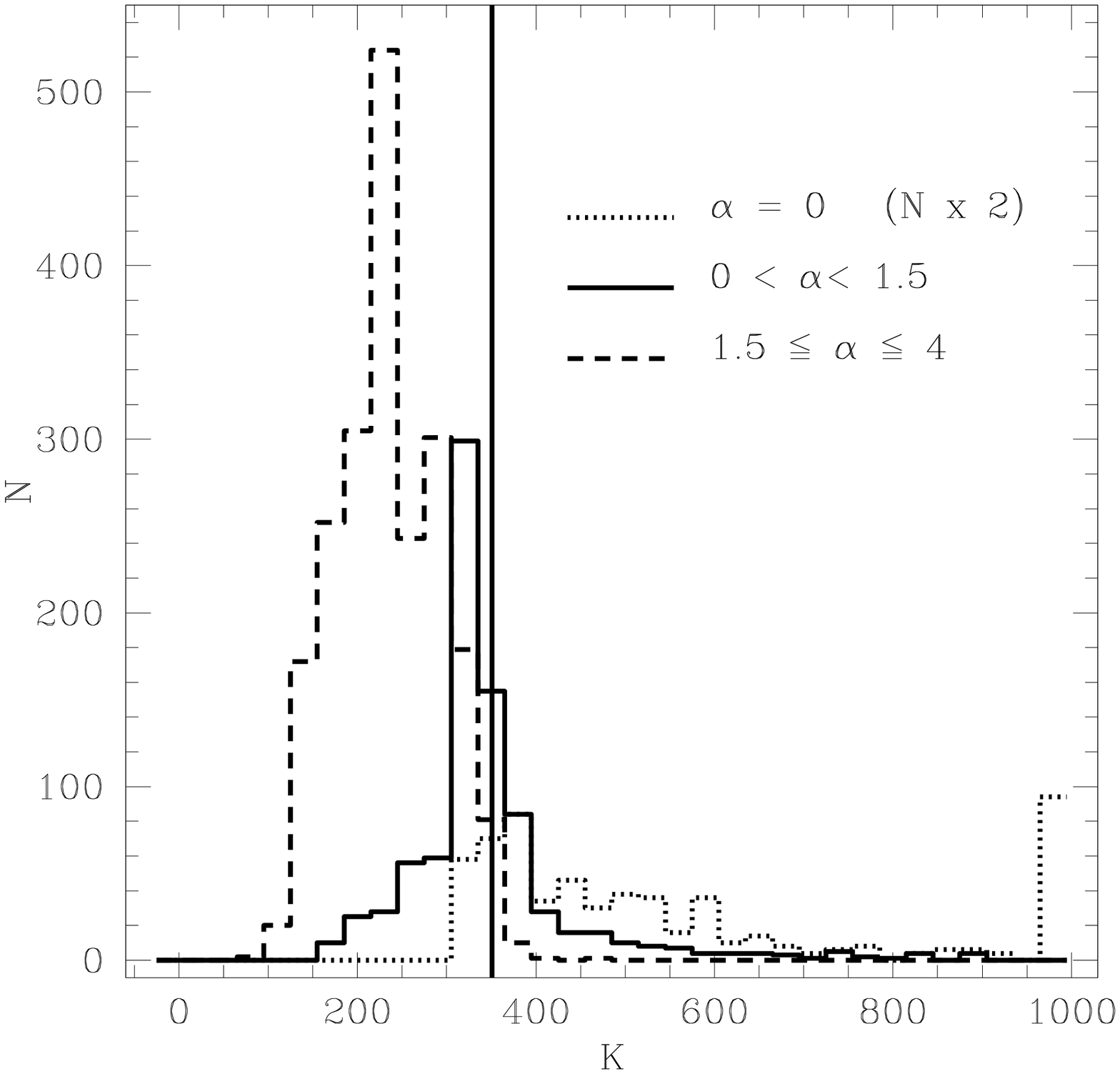}{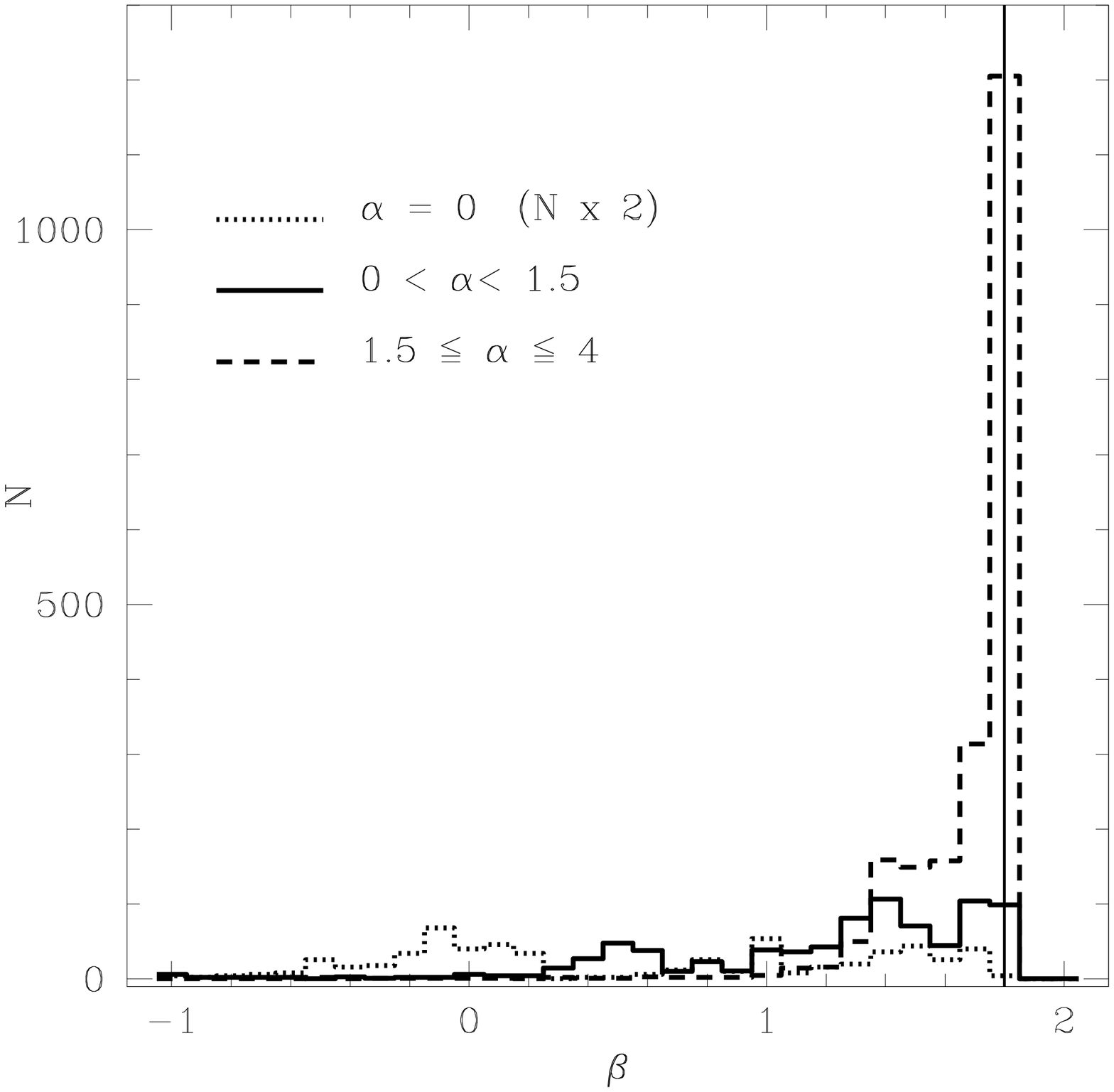}
\caption{The range of K and $\beta$ for the chosen
parameters in Table \ref{tab:2par} as a function of $\alpha$.  The vertical line marks the values of $K$ and $\beta_e$ in the Analytic Solution.  Histograms for $\alpha = 0$ have been multiplied by $2$ for clarity. 
\label{fig:Kbetaalpha}}
\end{figure}

Figure \ref{fig:Kbeta}  shows histograms of $K$ and $\beta$ derived from our
Fit Solution for our parameter space.  
While most models cluster around the Analytic
Solution, there is a spread that is dependent on some of the input
parameters.
In Figure \ref{fig:Kbetaalpha} we show how the values of 
$K$ and $\beta$ depend
on the $\alpha$ parameter.  Models that are far from the 
Analytic Solution (i.e. $K=1000$, $\beta = 0$) tend to have 
small values of $\alpha$.

\begin{figure}[!ht]
\epsscale {1.0}
\plotone{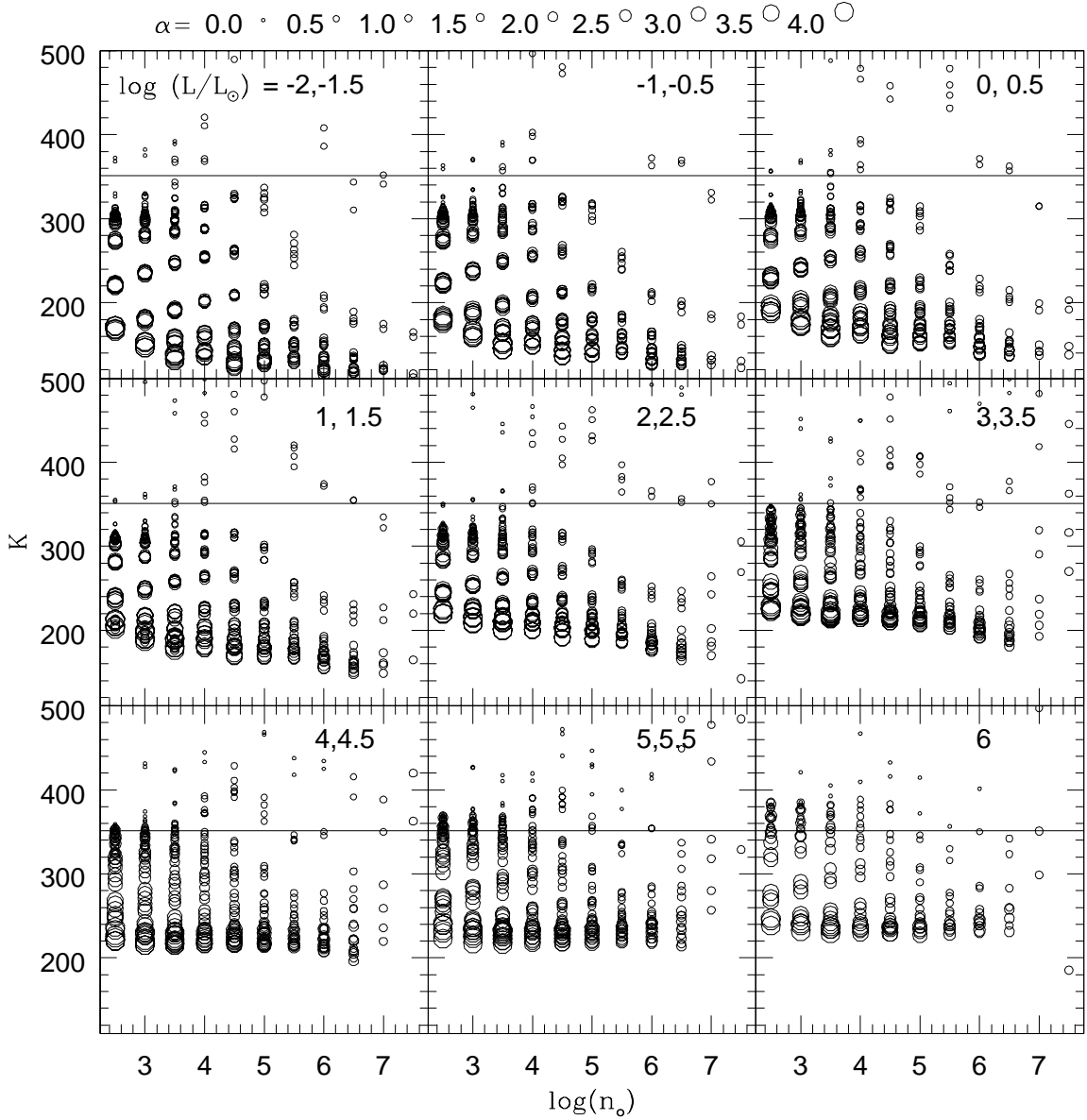}
\caption{The horizontal line marks the value of $K$ in the Analytic Solution. 
Individual models are plotted as circles of various sizes.  The size of the 
circle indicates the value of the $\alpha$ parameter as noted in the top of the
plot.  
The nine separate plots each show $K$ as a function of $n_o$ for nine
different luminosity regimes.  
In the top-left box, log $L/\Lsun$ is $-2$ or $-1.5$, 
as indicated at the top-right corner in the box.  
The bottom-right box shows models with the highest luminosities.  
\label{fig:Kall}}
\end{figure}

\begin{figure}[!ht]
\epsscale {1.0}
\plotone{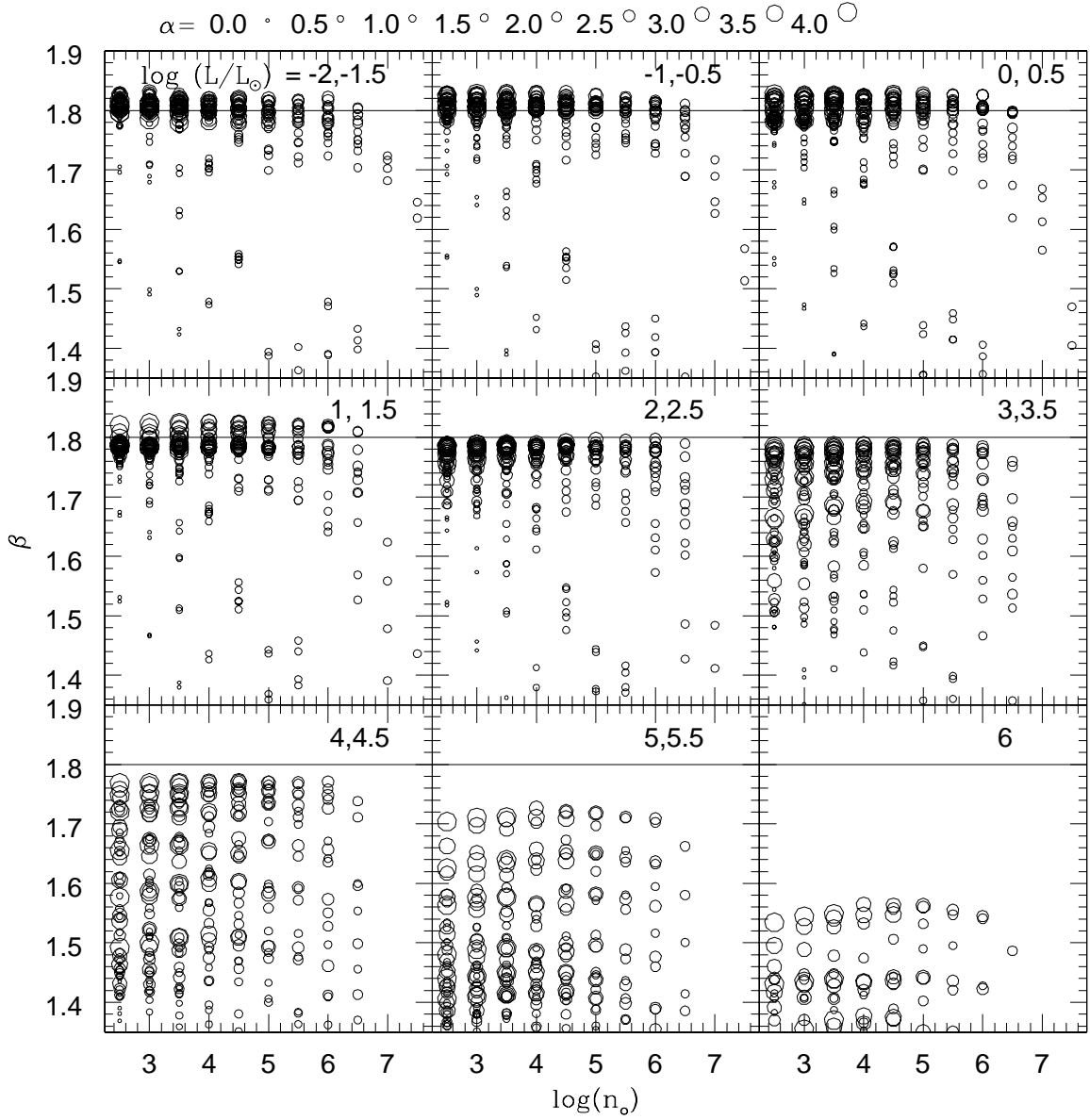}
\caption{The horizontal line marks the value of $\beta_e$ in the Analytic Solution.  Same plot details as Figure \ref{fig:Kall} except $\beta$ is 
plotted rather than $K$. 
\label{fig:betaall}}
\end{figure}

In Figures \ref{fig:Kall} and \ref{fig:betaall}, we show how different parameters
determine the values of $K$ and $\beta$.
Figure \ref{fig:Kall} shows that luminosity and the value of $K$ are 
positively correlated. 
Although we have explicitly 
removed the luminosity dependence from equation (\ref{eq:A}), there is still some
dependence of $K$ on luminosity.  
This can be understood in terms of radiative trapping.
For high luminosities, more photons at shorter
wavelengths exist farther from the star, which increases the size of the 
region of high optical depth and therefore increases the value of K.
Also, as $\alpha$ increases, $K$ decreases.  Increasing $\alpha$ shrinks the 
size of the region of high optical depth due to the buildup of material 
close to the star which absorbs the high energy photons.
Figure \ref{fig:betaall} shows that as $\alpha$ decreases and 
luminosity increases, 
$\beta$ moves farther from the Analytic Solution.  
The drastic drop of $\beta$ at high luminosities can be understood by
comparing the two models shown in Figure \ref{fig:fidL}.  
In the two models, the radius at which the dust temperature turns up 
(i.e., when the material changes from optically thin to optically thick)
moves out in radius as the luminosity increases.  
Since we model the value of $\beta$ using only the outer 
25\% of the material, we expect to be considering only the optically thin
material.  However, as the luminosity increases, the radius at which the
transition from optically thick to optically thin material moves outward. 
Therefore, at higher luminosities, our calculation of $\beta$ becomes
influenced by the optically thick region.  
This is the reason $\beta$ moves away from the optically thin Analytic
Solution as luminosity increases.

In order to calculate values of $\beta$ and $K$ not modeled in our parameter
space, we interpolate between known values in our look-up table.  
We use the method for interpolating in two or more dimensions described in 
Chapter 3 of \citet{numrec}.  
This method involves solving successive one-dimensional interpolations.
We modified the POLIN2 subroutine to interpolate in 4 dimensions.  
The actual method of interpolation that we used was the polynomial
interpolation method over 3 known quantities 
from the subroutine POLINT in \citet{numrec}. 
We tested our interpolation method by calculating extra models, not 
included in our model grid, and comparing the results between the 
real solution and the interpolated value.  The results are tabulated in Table
\ref{tab:interp}. 
We have varied all of the parameters by different amounts and found that the largest percent difference
from the model to the interpolated value is $6.0\%$.  The largest differences
occur at the extremes of the parameter-space at small $\alpha$ and large luminosity.  Large errors at small values of $\alpha$ are likely due to the large scatter of values of $K$ and $\beta$ at small $\alpha$ (see Figures \ref{fig:Kall} and \ref{fig:betaall}).  
The large differences at large luminosities are probably due to the rapid change of the value of $\beta$ with luminosity (see Figure \ref{fig:betaall}).  
Although $K$ does not change as much as $\beta$, 
the value of $K$ is dependent upon the value of $\beta$.

Based on the previous discussion in this section, we use
the Numerical Dust Temperature method
described in \S \ref{sec:2.2} to calculate the dust temperature 
in the remainder of the paper.  
Since we are primarily interested in the dust temperature far from 
the luminosity source, we find that we can model
the \emph{shape} and \emph{magnitude} of the 
dust temperature distribution most accurately with the Numerical Dust
Temperature method.  
There might be some error in the calculation of $K$ and $\beta$ 
for models with large luminosities or small $\alpha$'s.  
However, we don't expect very small values of $\alpha$ to be common 
\citep{mueller}.
Also, stars that will eventually have high luminosities are rare 
and will take awhile to reach this state, therefore we do not 
expect many stars to have high luminosities during the early
stages of star formation which we model.



\section{Gas Temperature Calculation}\label{sec:gas}
After the dust temperature as a function of distance from luminosity
sources is derived for positions near stars in a cluster using 
the look-up table, the gas temperature can be calculated
assuming gas energetics balance. 
We calculate the gas temperature using a  
gas-dust energetics code 
which includes energy transfer between gas and dust via collisions, 
heating by cosmic rays, 
and molecular cooling (see \citet{doty}
and the appendix in \citet{young} for a more detailed description).
We assume that the dust to gas mass ratio is $\eta_{dg}=4.86\ee{-3}$ 
\citep{hm}
and the grain cross-section per baryon is $6.09\ee{-22}$
\citep{young}.
The cosmic ray ionization rate is $3.00\ee{-17}$ s$^{-1}$ \citep{vans}
and the energy deposited per cosmic ray ionization is $2.00\ee{1}$ eV 
\citep{goldsmith} in our models.
We take the fractional abundance of CO relative to H$_2$ to be $1.0\ee{-4}$ 
from Figure 8a of \citet{lee}.
We assume that the region we are studying is deep within a larger molecular cloud.
Therefore, there is no interstellar radiation field impinging on the outer
bounds of the cloud and the photoelectric effect on PAH's is 
not present.

The model dependent input parameters are
T$_{dust}$, local density, column density, and local velocity dispersion (b).  
T$_{dust}$ is calculated with the procedure described in \S \ref{sec:2.2}.
Local density and column density can be derived from our
input density profile.  
The column density is calculated radially from the point of interest to the edge of the 
system.  The edge of the system is defined as either the point at which the density is lowest or some fiducial value (as discussed later in \S \ref{sec:multiple}).
The velocity-spread parameter, $b$, is defined for a Maxwellian velocity 
distribution as $b=(2kT/m)^{1/2}$ \citep{spitzer} 
and is assumed to be 1 km/s throughout this paper. 



\begin{figure}[!ht]
\epsscale{1.}
\plotone{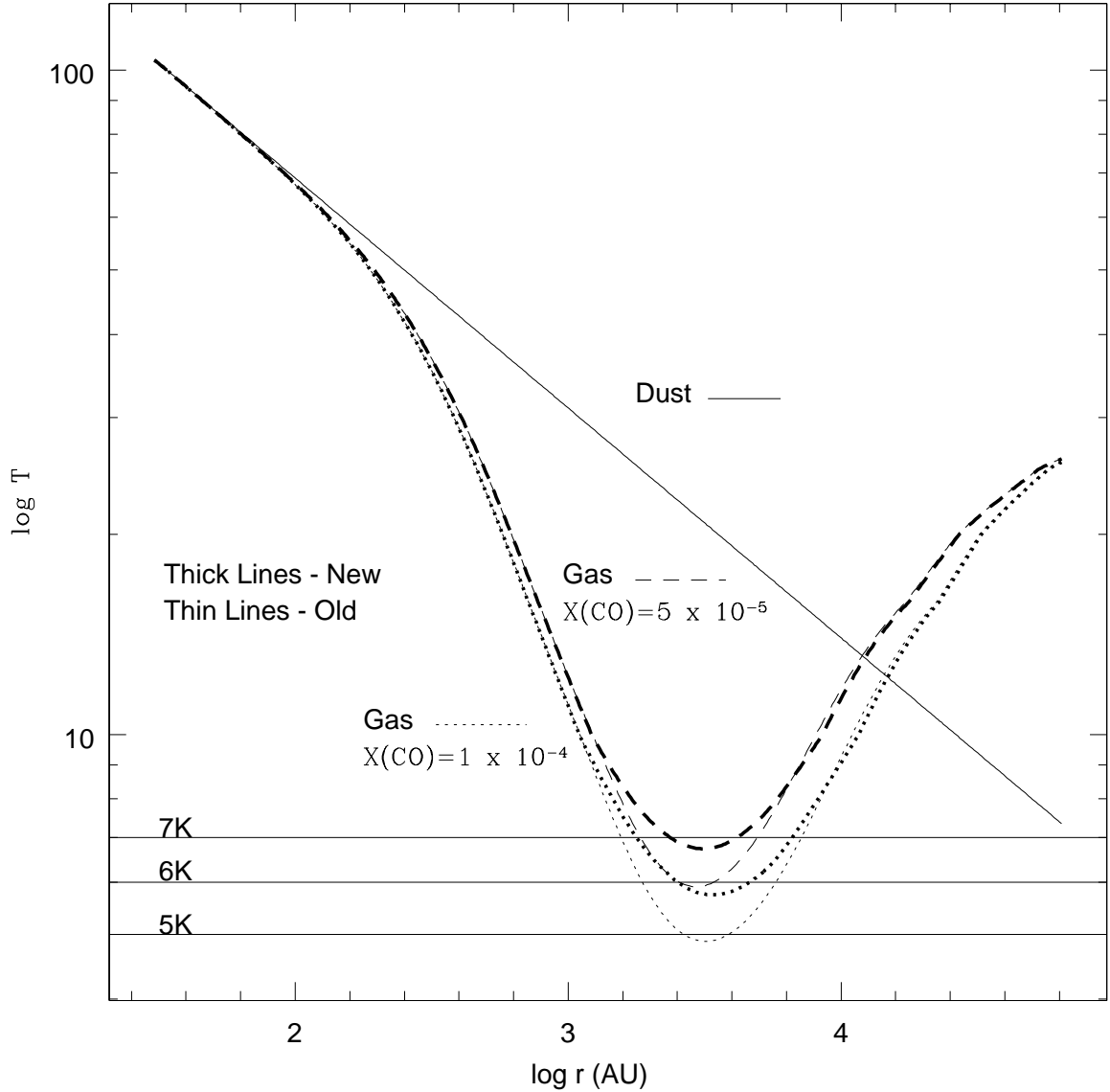}
\caption{Gas temperature distribution and comparison of old and new cooling rates.  
Solid line shows Fit Solution to dust temperature.
The model parameters are log ($L/\Lsun=1$), log $(r_{out}/1pc)=-0.5$, 
log $n_o= 4.5$, and $\alpha=2.5$. 
Dashed and dotted line show gas temperature for X(CO)=$1\ee{-4}$ and 
X(CO)=$5\ee{-5}$, respectively.
Old and new cooling rates are shown in the thin and thick lines, respectively.
\label{fig:compare}}
\end{figure}

Figure \ref{fig:compare} shows the variation of gas temperature with distance
from a stellar heating source for two values of X(CO).
Close to the source, the dust and gas temperature are coupled due to 
collisional interactions of the dust with the gas.  
As the density decreases, collisions between the dust and gas become less
frequent and the gas is able to cool via molecular (mainly CO)
rotational transitions.  
Then as the density continues to drop
and there is less CO to cool the gas, cosmic
ray heating becomes the dominant heating source
and the gas temperature increases.
For a gas with less CO (X(CO)$=5\ee{-5}$), the cooling is not as efficient
and the temperature is larger.
As various other parameters change in our models, 
different heating and cooling 
terms dominate and the minimum and maximum temperatures vary.  
We discuss this in more detail in the following section 
(\S \ref{sec:multiple}).

As seen in Figure \ref{fig:compare},
the gas temperature falls below $10$K
when gas collisions are the dominant cooling method.  
Our gas cooling rate calculations are based on the work of \citet{nk93} and 
\citet{nlm95} which only extend down to 10K because the \hh-CO collisional
rates were not defined below 10K at the time of their work.  
We expect the cooling rate to drop drastically as the temperature 
approaches zero and we have attempted to adjust our cooling rate calculation
to account for this.
Our first attempt to calculate the cooling rate ($\Lambda$) below 10K 
involved calculating the rate of change of log $\Lambda$ with log $T$ 
between 20K and 10K ($d$log$\Lambda$/$d$log$T$) and 
applying it to temperatures below 10K.  
This method has been used in \citet{young} successfully 
when dust-gas collisions, rather than \hh-CO collisions, 
are the dominant mode of energy transfer at low temperatures. 
However, for our models, \hh-CO collisions dominate at low temperatures 
and to take this into account we modify the method of 
calculating $d$log$\Lambda$/$d$log$T$ using the 
Large Velocity Gradient model (LVG).
Instead of extrapolating the cooling rate to low temperatures,
we extrapolate the CO rate coefficients from \citet{fl} from 10K to 5K.
Then we calculate $d$log$\Lambda$/$d$log$T$ between 10K and 5K 
using the LVG model.
We apply the new values of $d$log$\Lambda$/$d$log$T$ between 10K and 5K 
to our gas energetics model.  
In Figure \ref{fig:compare} 
we compare the temperature derived using the old and new 
method of calculating the gas temperature.  Our change increases
the temperature by approximately 1K at the minimum value.

\section{Gas and Dust Temperature with Multiple Sources}\label{sec:multiple}

\subsection{Two Sources}
In order to calculate the gas temperature between two sources, we must 
first calculate the dust temperature.  We do this using the formalism
discussed in \S \ref{sec:2.2}.  Once we have determined the dust 
temperature, we can use our energetics algorithm to calculate the 
gas temperature.  
Around each source we place a density profile.  In order for this 
to be realistic, we choose a density, $n_{eq}$ at which we have the 
two density profiles meet.  
The value of $n_{eq}$ sets the distance between the sources, i.e.  
smaller values of $n_{eq}$ place the sources farther apart.

\begin{figure}[!ht]
\epsscale{1.0}
\plotone{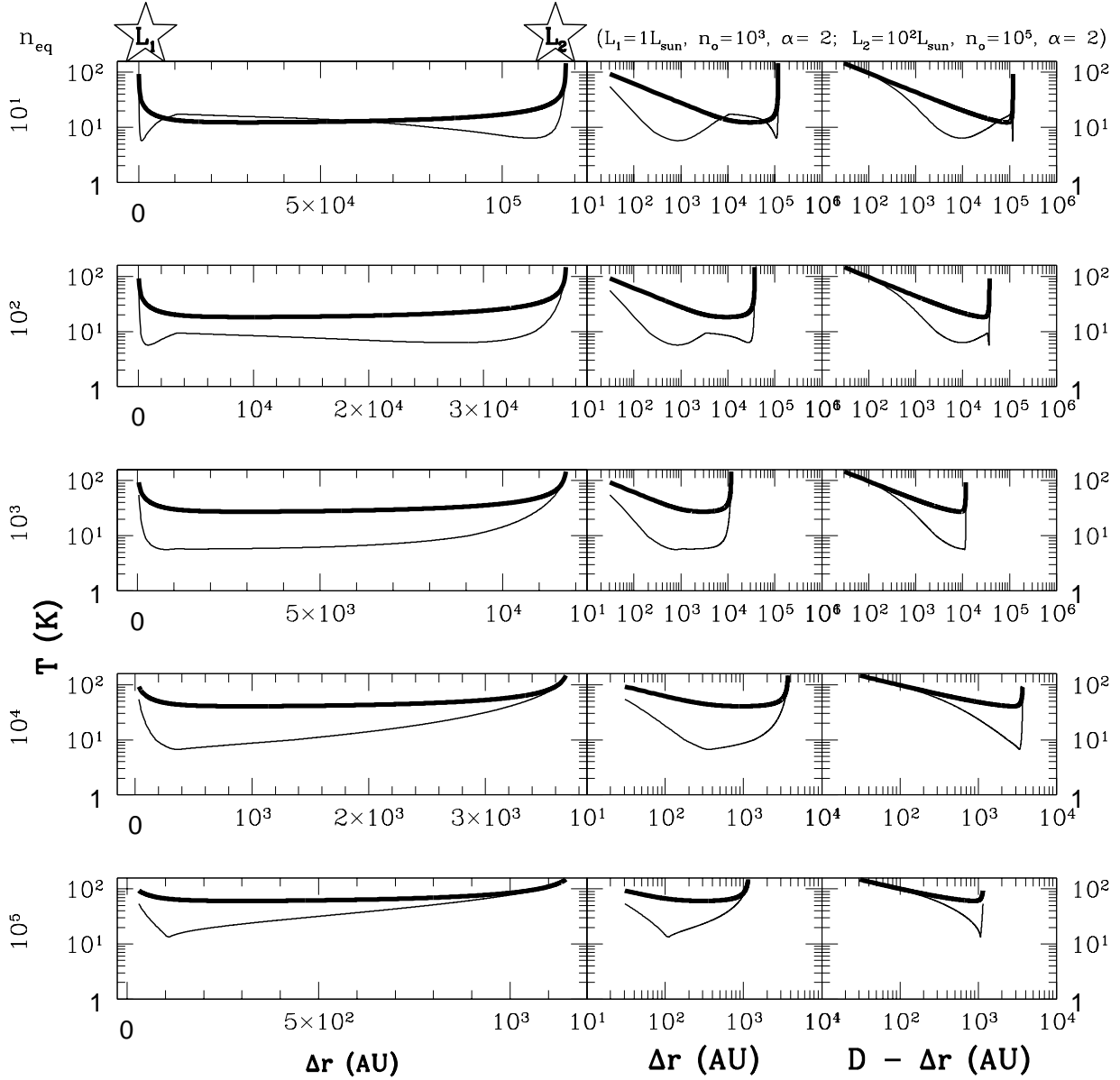}
\caption{Gas temperature distribution between two sources.
Plots show dust (thick line) and gas (thin line)
temperature as a function of distance between two stellar 
sources.  
From top to bottom, distance between sources is decreasing,
such that the gas density surrounding the two sources
agrees with the value, $n_{eq}$ quoted on the left.
The source on the left (1) has the parameters $L=1\Lsun$, $n_o=10^3$,
and $\alpha=2$.  
The right source (2) has $L=10^2\Lsun$, $n_o=10^5$, and $\alpha=2$.
The three horizontal plots for the different values of $n_{eq}$
show the same data from different perspectives.  The large plot 
on the left is plotted on a linear scale from Source 1.  The two smaller
plots on the right are plotted on a logarithmic scale, from the 
perspective of Source 1 (middle) and Source 2 (right).  
This is done to show structure close to the individual sources.
\label{fig:two}}
\end{figure}

Figure \ref{fig:two} shows the dust and gas temperature profile for
increasing values of $n_{eq}$, i.e. smaller separations. 
An interesting feature of this plot is that if we only look at the region
between the two sources, we find a large variation in gas temperature.
This is due to the higher densities sampled as the sources move
closer together. 
The top panel shows sources that are far apart and we see that the 
maximum gas temperature between the two sources is $\sim 20$K and 
the minimum gas temperature is $\sim 7$K.  
Cosmic ray heating, though relatively weak, 
can warm the material sufficiently far from luminous sources.
As the sources move closer together, cosmic rays become less important
until the temperature between the sources ceases to increase, 
whereas dust heating becomes more important and the 
gas is not able to cool efficiently and the minimum temperature
between the sources rises.

\subsection{Three Sources}\label{sec:three}
\begin{figure}[!ht]
\epsscale{1.1}
\plottwo{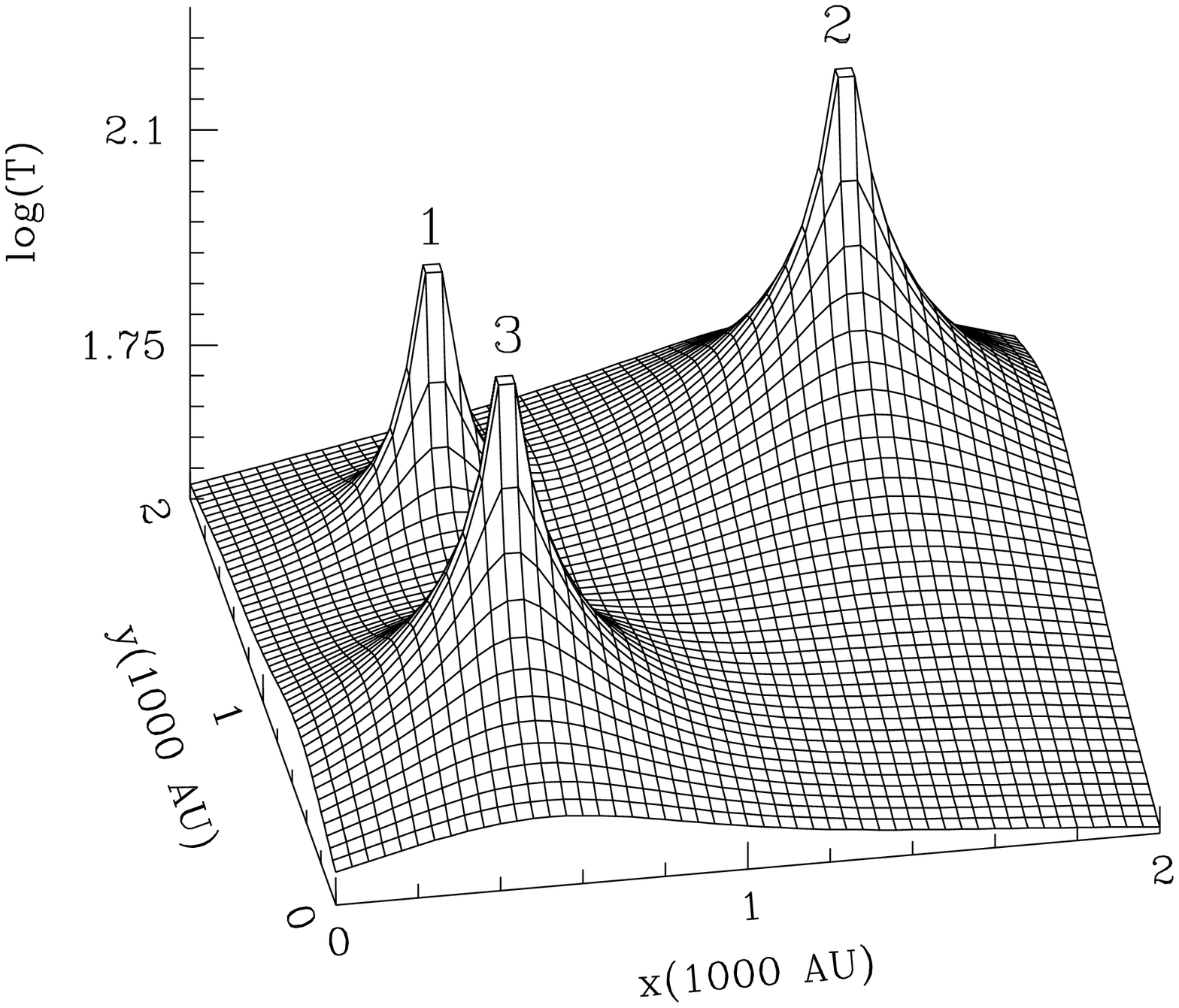}{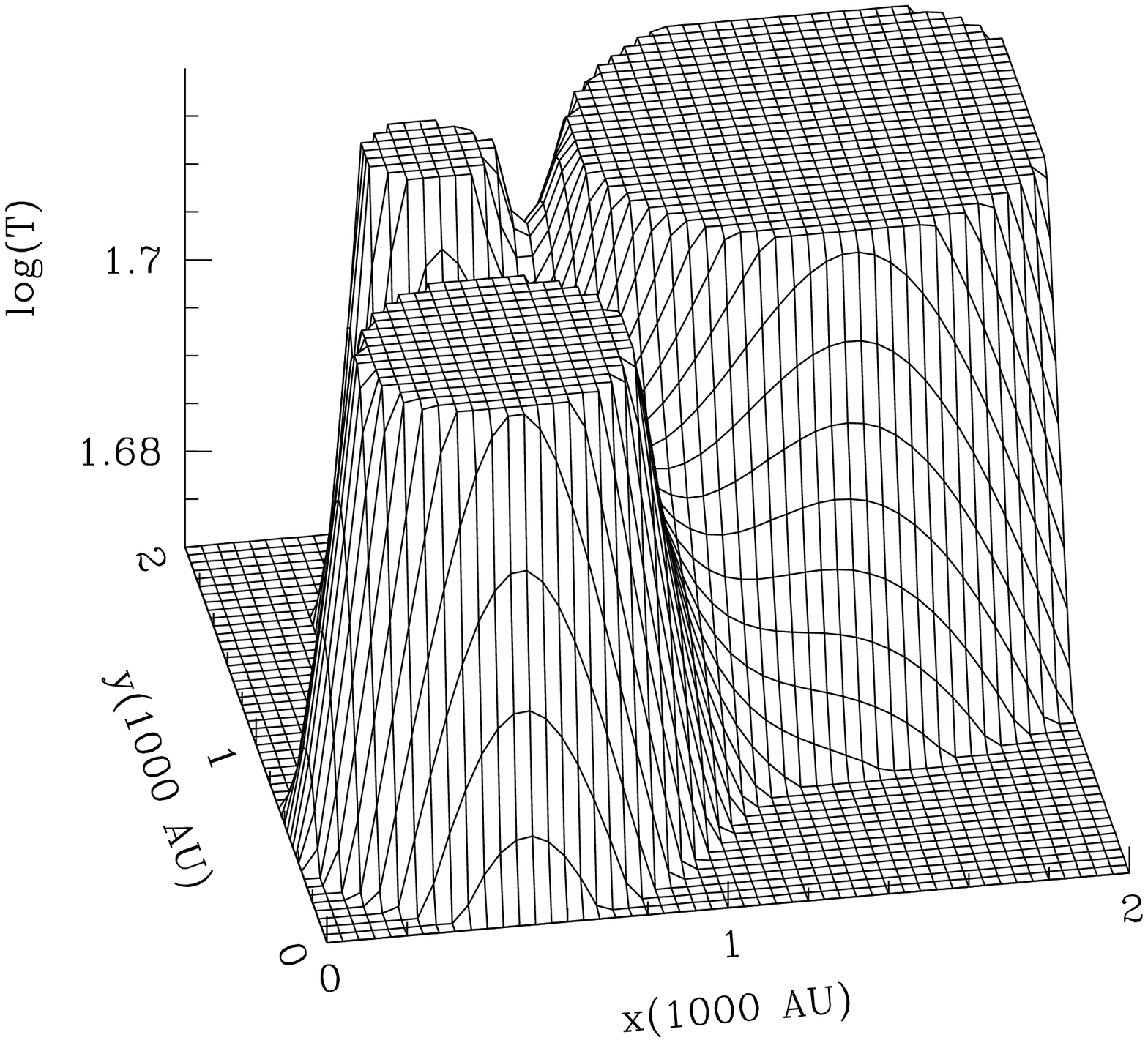}
\caption{Surface plots of dust temperature.  Sources are labeled according to the parameters listed in Table \ref{tab:source}.  The first figure shows the full range of dust temperatures and the second figure shows a smaller range of dust temperatures.
\label{fig:tdust3}}
\end{figure}

Here we calculate the gas and dust temperature distribution around three sources.
The three sources were placed on three of the 
corners of a square with sides of length 1000AU.  
The least luminous source was placed on the corner between the other two
sources. 
The parameters assumed for each source are given in Table
 \ref{tab:source} as well as the position of the sources in a 2000AU x 2000AU grid.
We calculate the dust temperature using the method 
described in \S \ref{sec:2.2} and show the results in Figure \ref{fig:tdust3}.

In order to determine the gas temperature, we first calculate 
the density ($\rho$) and column density ($N_{col}$) 
at each point in the grid due to all three sources, individually.
Then, at each point we choose the source which gives
the highest value of $N_{col}$ and use that source to calculate $N_{col}$, 
$\rho$, and the gas temperature.

\begin{figure}[!ht]
\epsscale{1.1}
\plotone{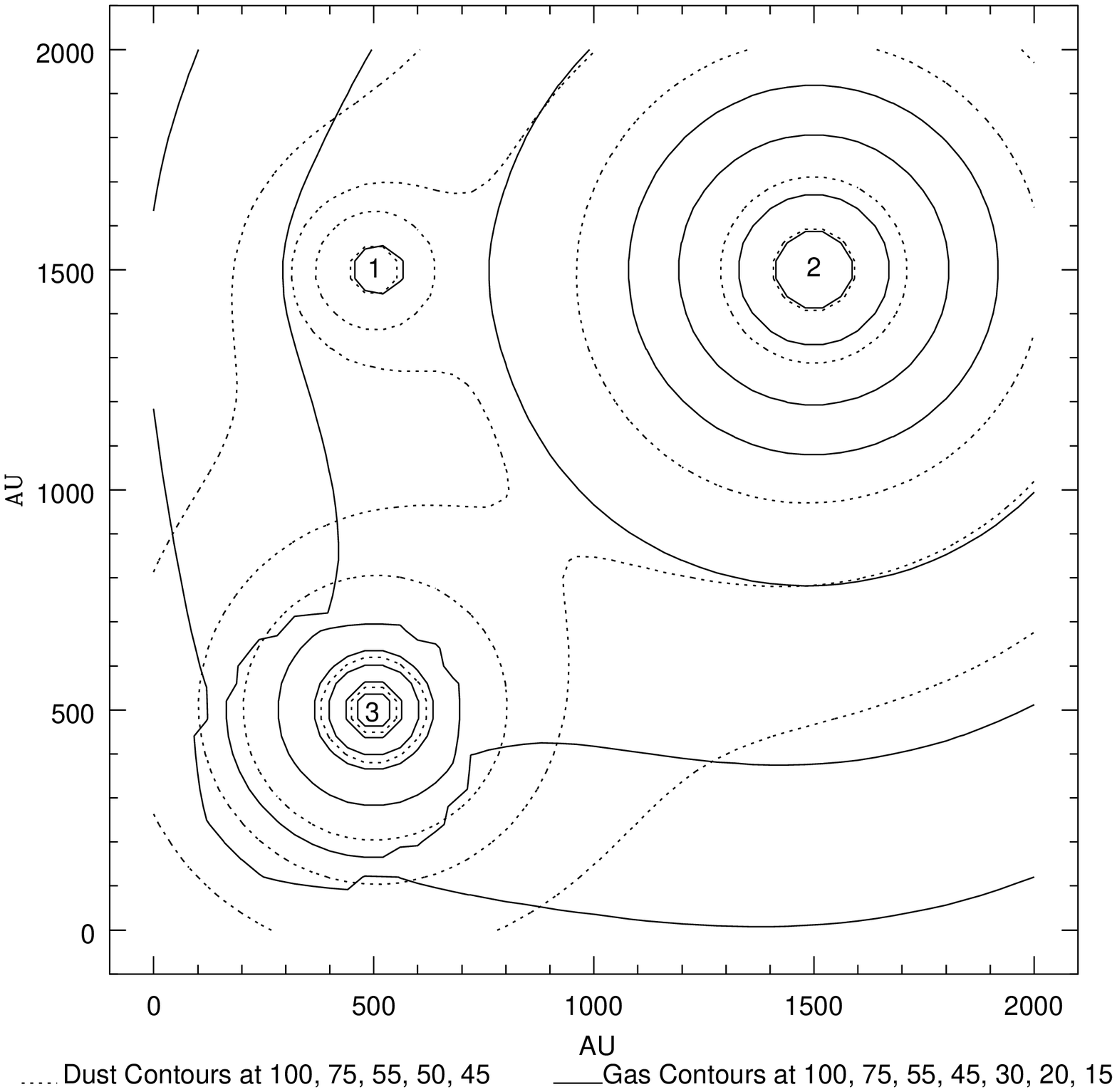}
\caption{Contour plots of dust and gas temperature using the Length of Square Method.
Sources are labeled according to the parameters listed in Table \ref{tab:source}.
\label{fig:td32000con}}
\end{figure}

\begin{figure}[!ht]
\epsscale{1.1}
\plottwo{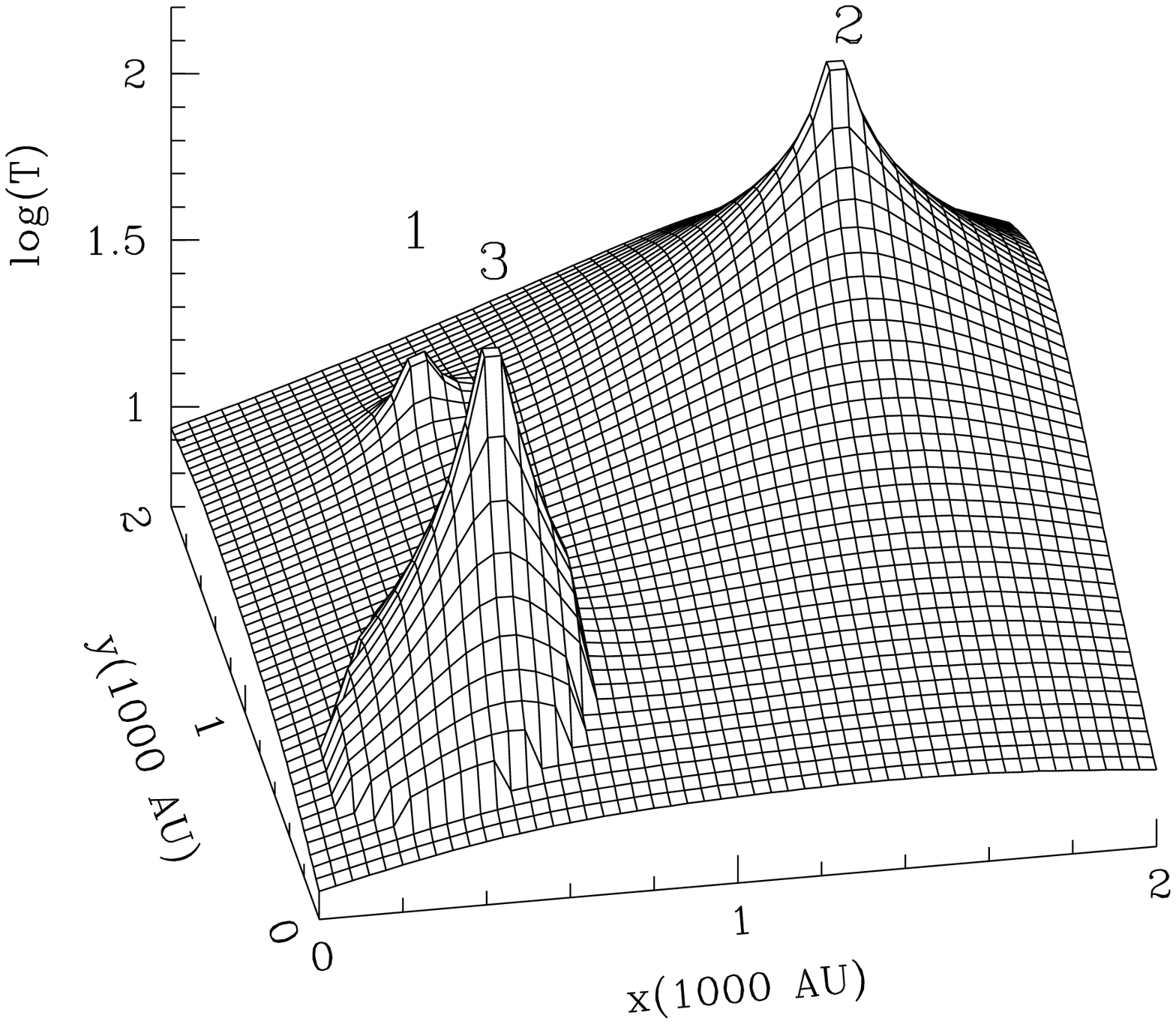}{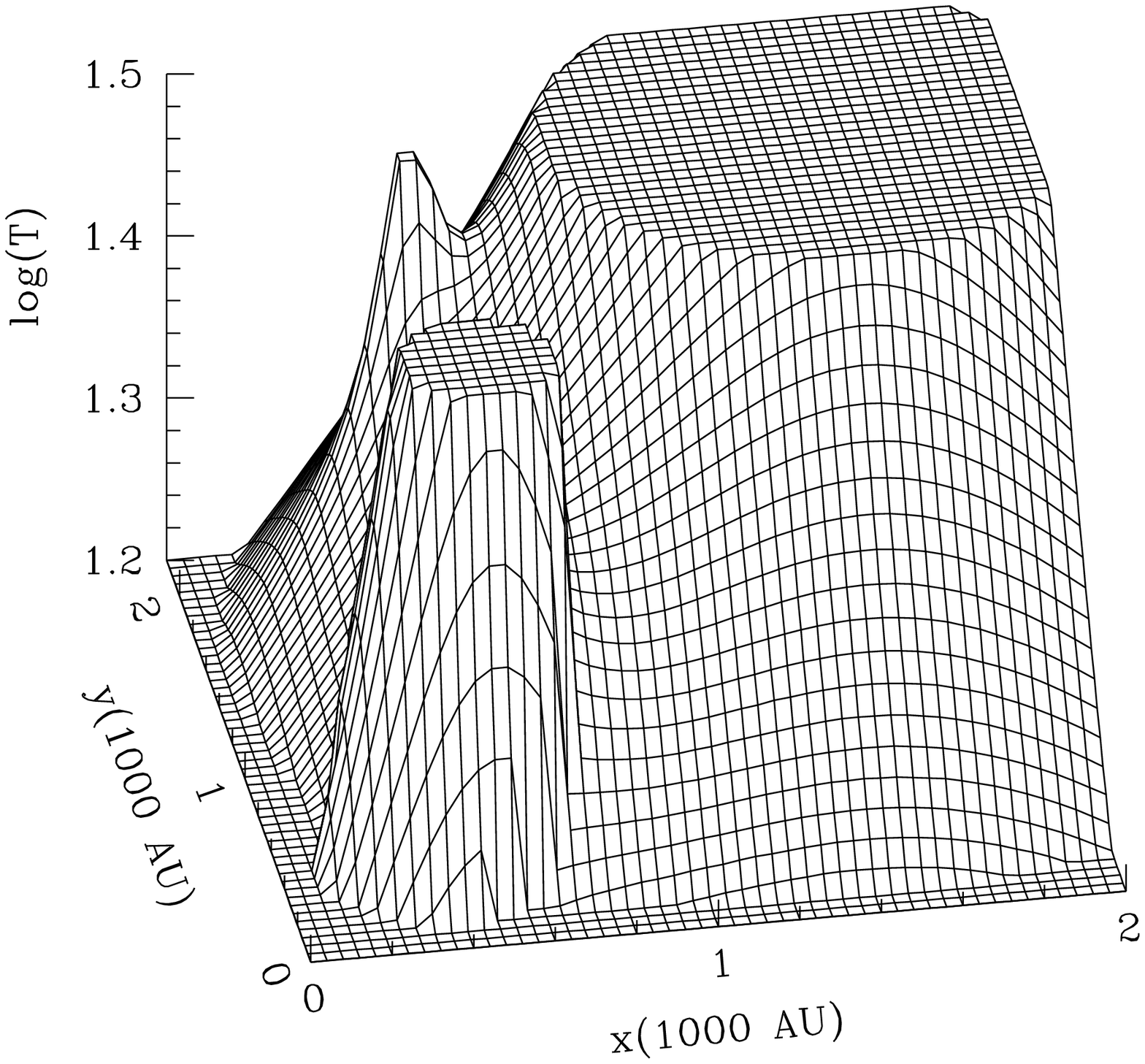}
\caption{Surface plots of gas temperature using the Length of Square Method. 
The same data is shown in both plots with a smaller range of gas temperature 
shown in the second plot.
\label{fig:td32000surf}}
\end{figure}

\begin{figure}[!ht]
\epsscale{1.0}
\plotone{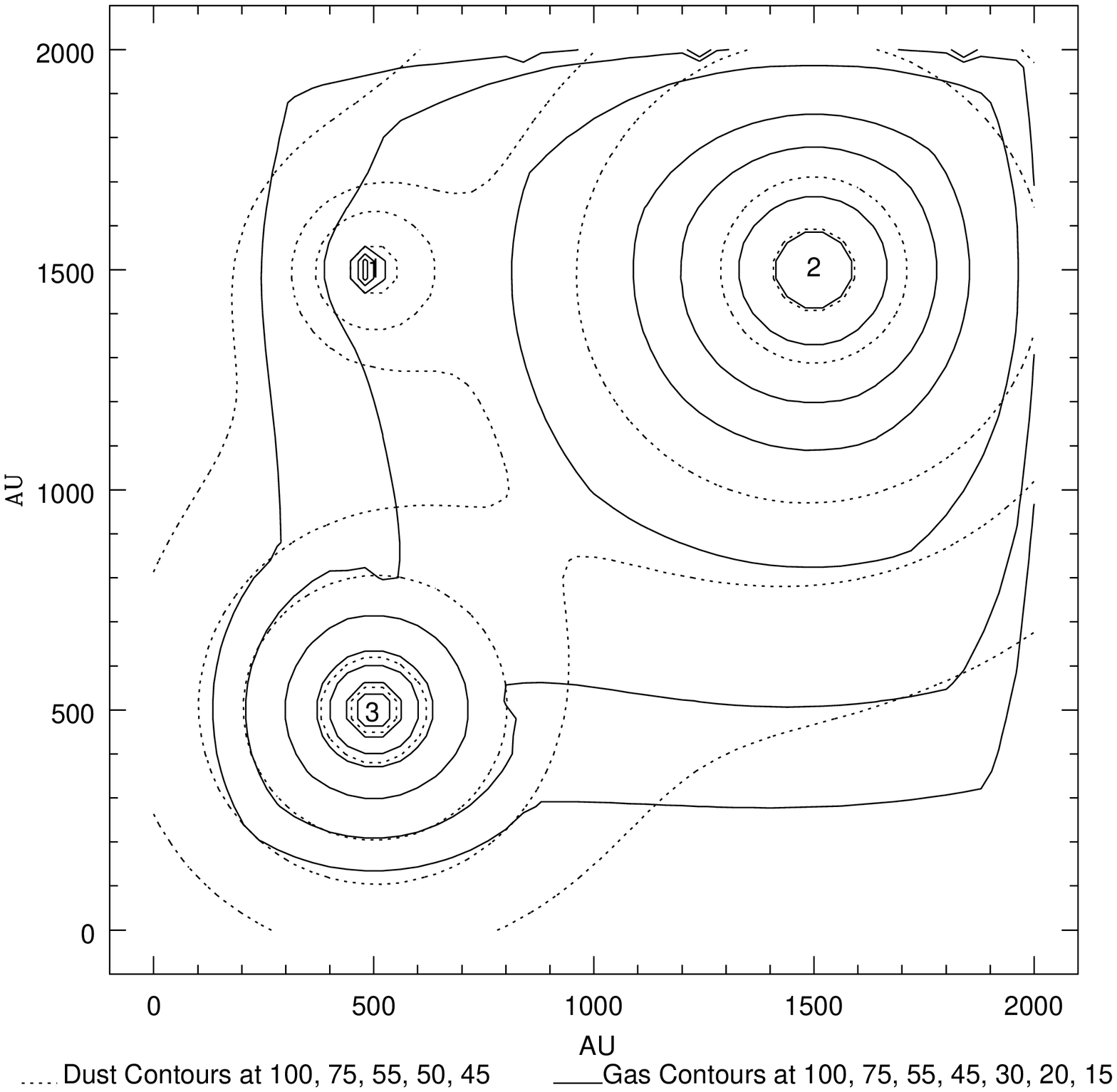}
\caption{Contour plots of dust and gas temperature using the Edge of Square Method.  
Sources are labeled according to the parameters listed in Table \ref{tab:source}.
\label{fig:td3econ}}
\end{figure}

\begin{figure}[!ht]
\epsscale{1.1}
\plottwo{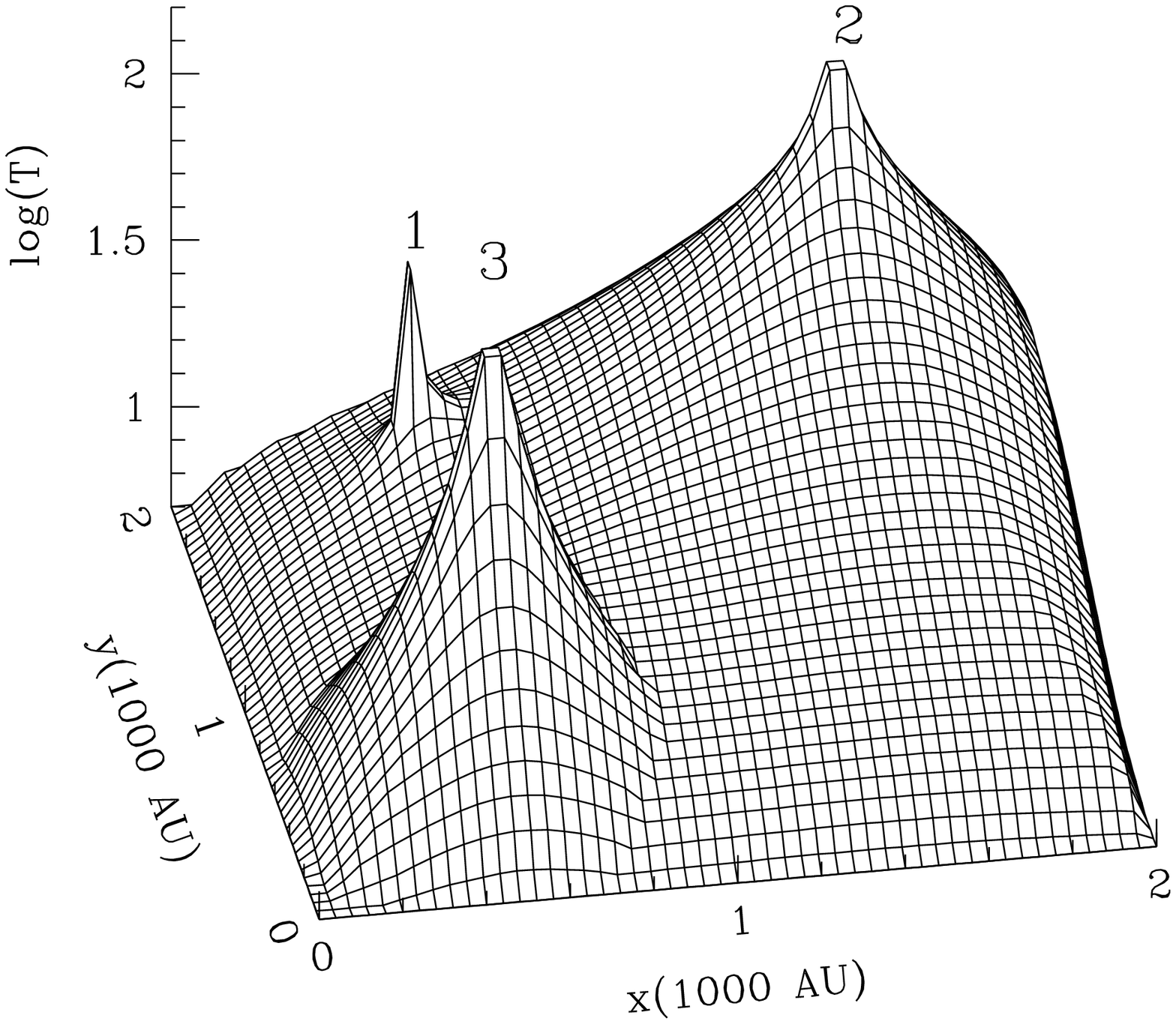}{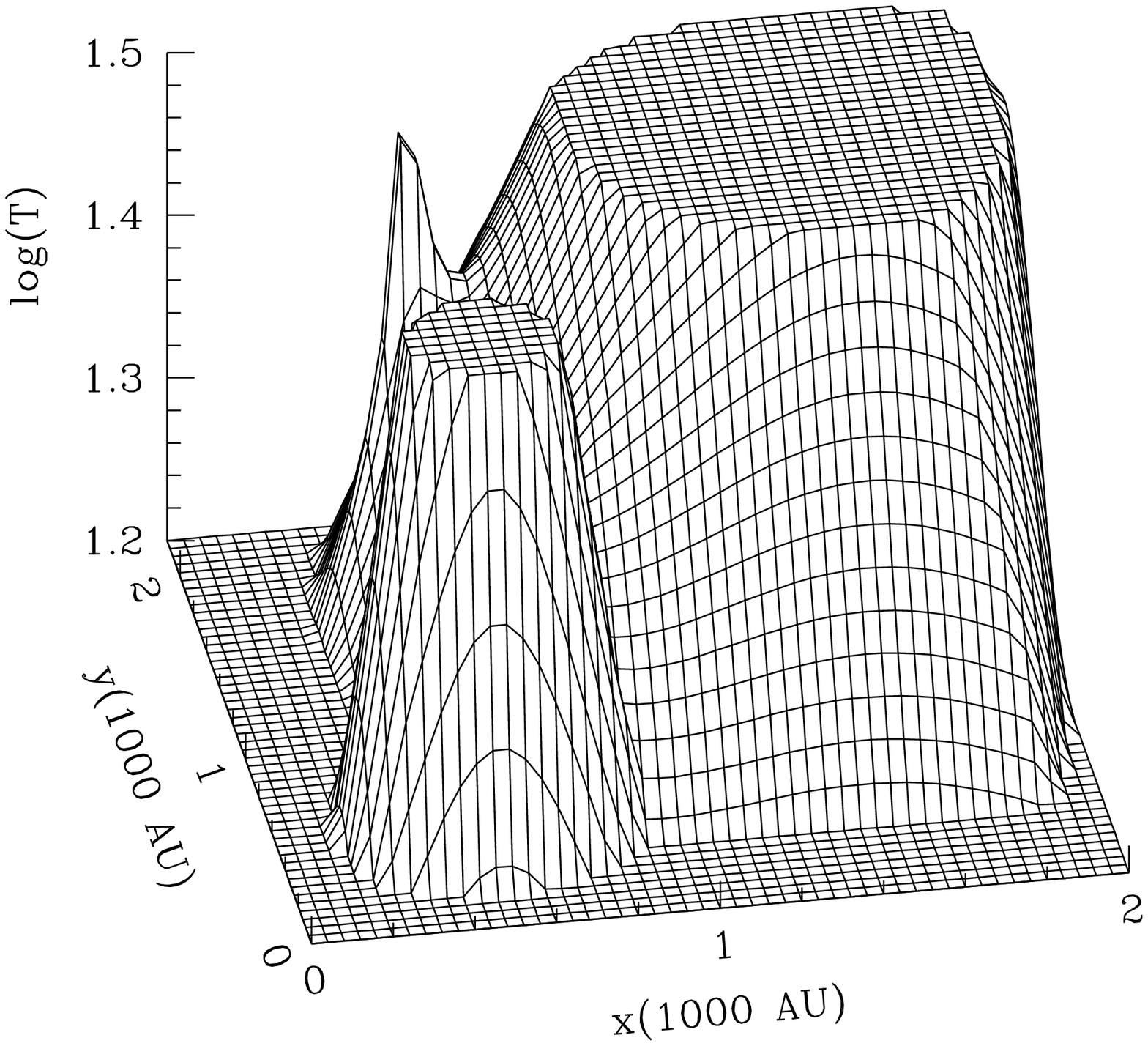}
\caption{Surface plots of gas temperature using the Edge of Square Method.
The same data is shown in both plots with a smaller range of gas temperature 
shown in the second plot.
\label{fig:td3esurf}}
\end{figure}

In order to calculate $N_{col}$, 
we integrate from the point of interest to the ``edge'' 
in the direction radially away from the source. 
We tested two methods of defining the ``edge.'' 
Our first method, the ``Length of Square Method,''
integrates from the point of interest to 2000AU from the source 
(Figure \ref{fig:td32000con} and \ref{fig:td32000surf}).
The value of 2000 AU was chosen to equal the length of the side of the square 
in which we place our sources.
Our second method, the ``Edge of Square Method,'' 
integrates from the point of interest to the edge of the region studied 
(Figure \ref{fig:td3econ} and \ref{fig:td3esurf}).
Using this method, the integration length depends on the direction
of integration. 
In both cases, we find that the gas and dust
temperature are not equal, unless we are in a 
region of high density close to a luminosity
source as seen in Figures \ref{fig:td32000con} and 
\ref{fig:td3econ}.  In these high density regions, the dust and 
gas temperature are coupled through collisions.  
As the density decreases, the gas temperature drops rapidly, 
compared to the dust temperature, due to the ability of the gas 
to cool through molecular transitions.
Another interesting feature of these plots is the detectability 
of Source 1, the dimmest object in the region, 
even though it is close to Source 2, the brightest source
in the region.

Two differences between the two methods of calculating $N_{col}$
are evident in Figures \ref{fig:tdust3} - \ref{fig:td3econ}.
The first is the the square shape of the contours near the edges 
when using the Edge of Square Method in Figure \ref{fig:td3econ}.  
This is an artifact of a square region of interest.
The second difference is the gas temperature near Source 1.  
It appears to be shifted in the direction of Source 2 for the Length of Square Method 
(Figure \ref{fig:td32000con}) and 
is shifted in the opposite direction for the Edge of Square Method 
(Figure \ref{fig:td3econ}).
For both methods, the dust temperature and gas-dust collisional 
heating is the same in this region.  
Therefore, the molecular cooling rate 
and the column density must be different.  
For the Length of Square Method the calculation of $N_{col}$ is dominated by 
Source 2 near Source 1.  
This is the reason the gas temperature contour around Source 1 is shifted 
toward Source 2.  
The value of $N_{col}$ drops as we move across Source 1.  This causes
a shift in the gas temperature toward Source 2 even though the 
dust temperature on either side of Source 1 is symmetric.
For the Edge of Square Method, Source 1 is able to increase the value of $N_{col}$ 
on the side facing away from Source 2 
over the value calculated from Source 2.  
This increase in $N_{col}$ causes the gas temperature to increase as well. 
Although neither of these methods is entirely correct, 
we use the Length of Square Method (adjusted for the size of the square) in the following
figures due to the square edge effects of the Edge of Square Method. 

\begin{figure}
\epsscale{1.25}
\plottwo{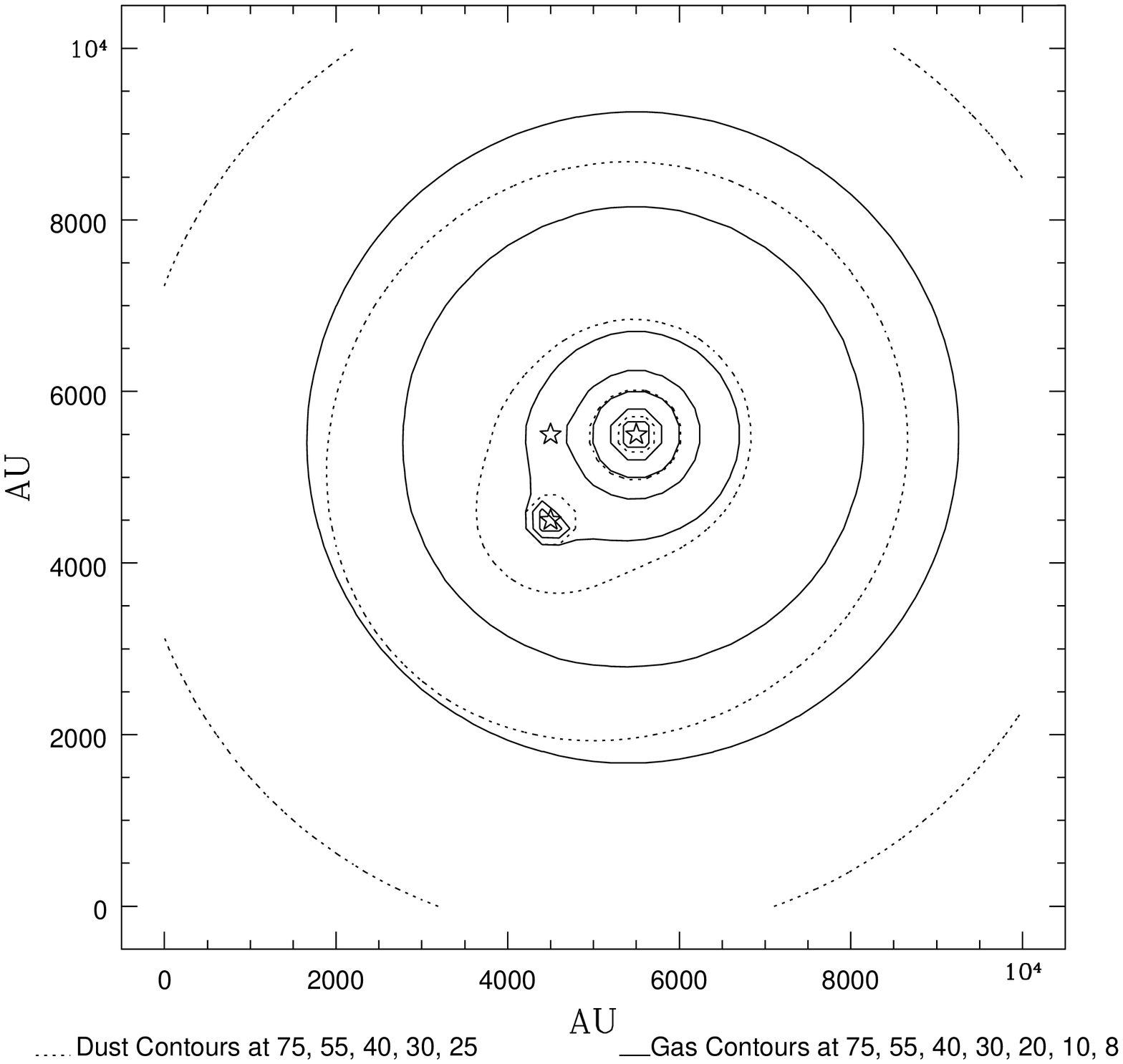}{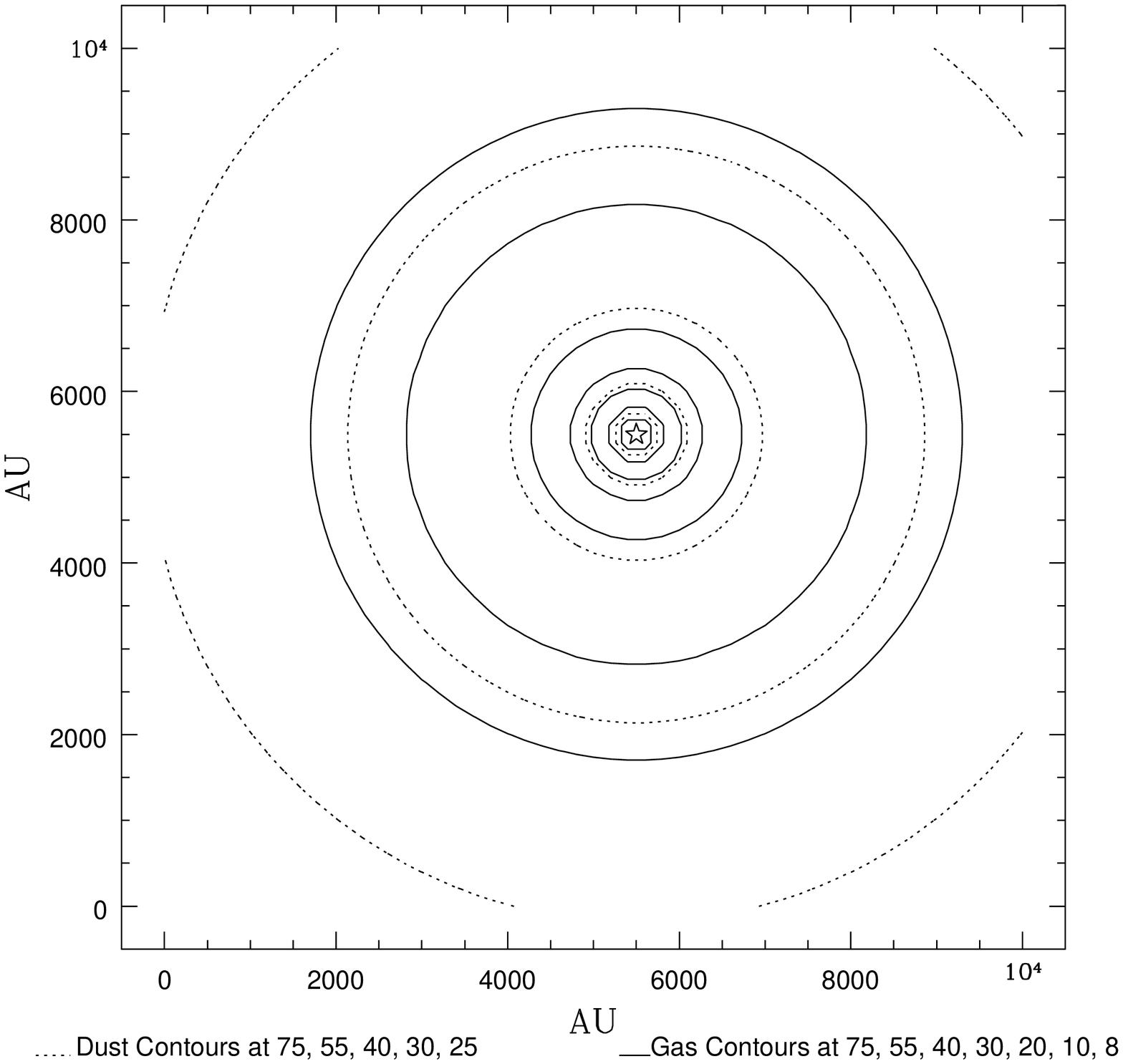}
\caption{Contour plots of dust and gas temperature with three sources and one source. 
\label{fig:z31}}
\end{figure}

In Figure \ref{fig:z31}, we compare the gas and dust temperature far from the sources
when we replace the three sources 
with a single source at the location of Source 2.  
We use the same 
parameters as Source 2 for our single source, however we have increased its
luminosity to 130$\Lsun$.  This figure illustrates the difficulty in 
determining the number of sources responsible for heating the gas and dust and
highlights the need for adequate spatial resolution.

\begin{figure}
\epsscale{1.1}
\plotone{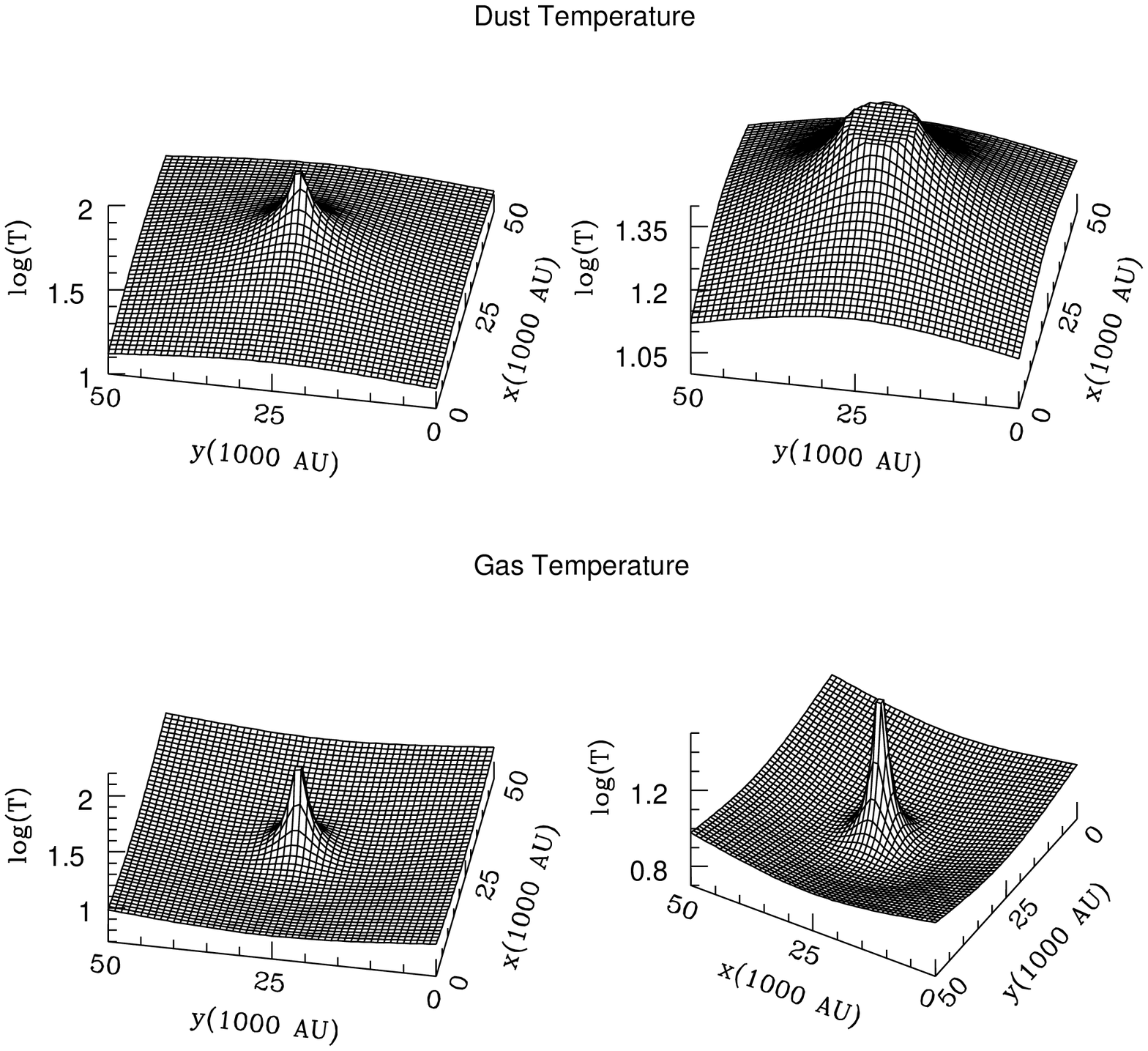}
\caption{Surface plots of dust and gas temperature. Top plots show dust temperature.  Bottom plots show gas temperature.
Plot on top left show complete sample of dust temperatures.  Top right plot
shows a smaller range in dust temperatures.   
Bottom right and bottom left show same data from two different orientations. 
Area sampled is 50000 x 50000 AU.  The Length of Square Method is used here.
\label{fig:zoomout}}
\end{figure}

In Figure \ref{fig:zoomout}, we have zoomed out of the region of interest.  
Notice that the dust temperature steadily decreases as the distance from the central three sources increases.  
Yet, the gas temperature slowly begins to rise. 
This is due to the decrease of effectiveness of CO cooling and the increase in 
heating by cosmic rays as the density decreases. 

\section{Conclusion}
\begin{figure}
\plottwo{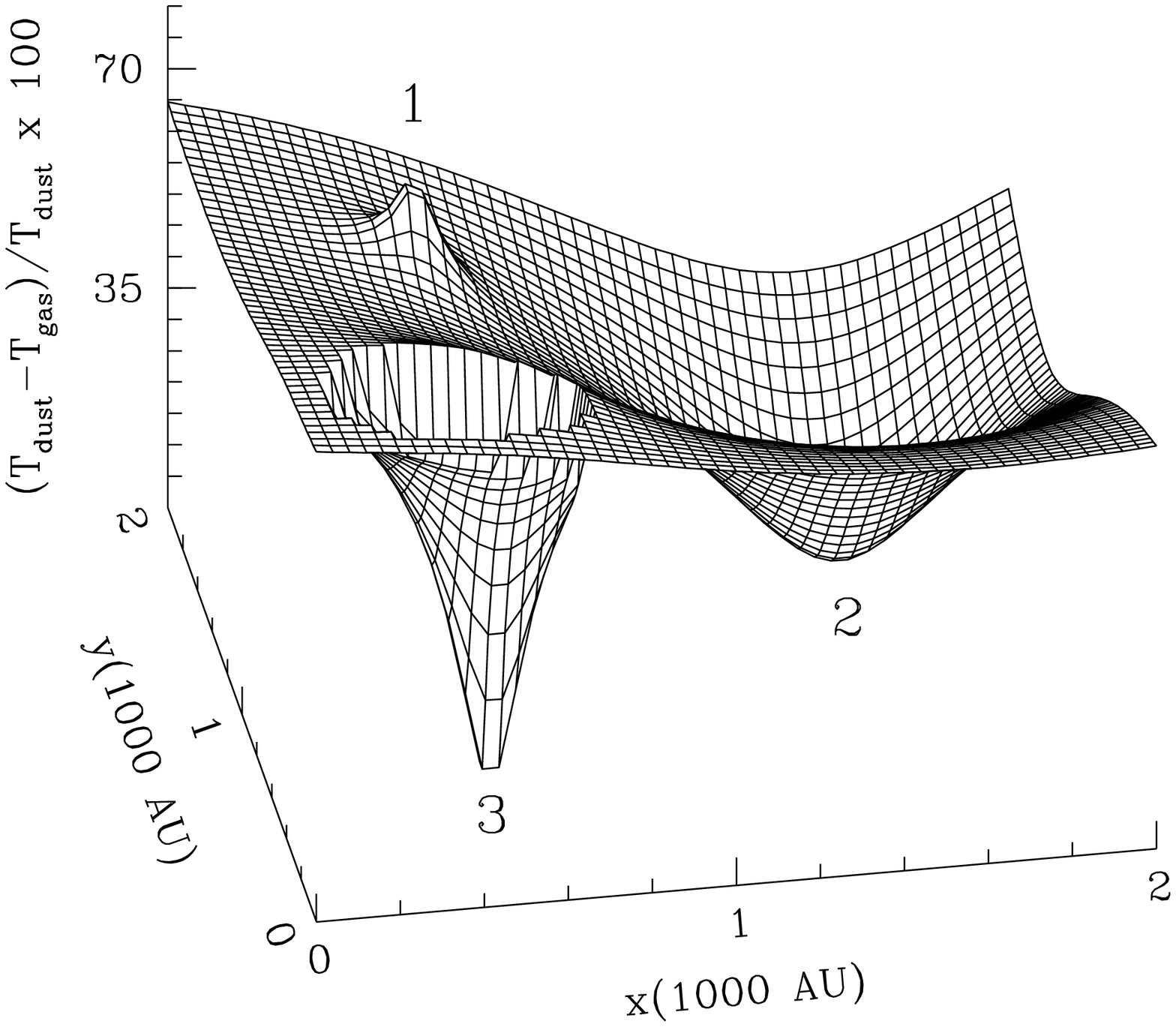}{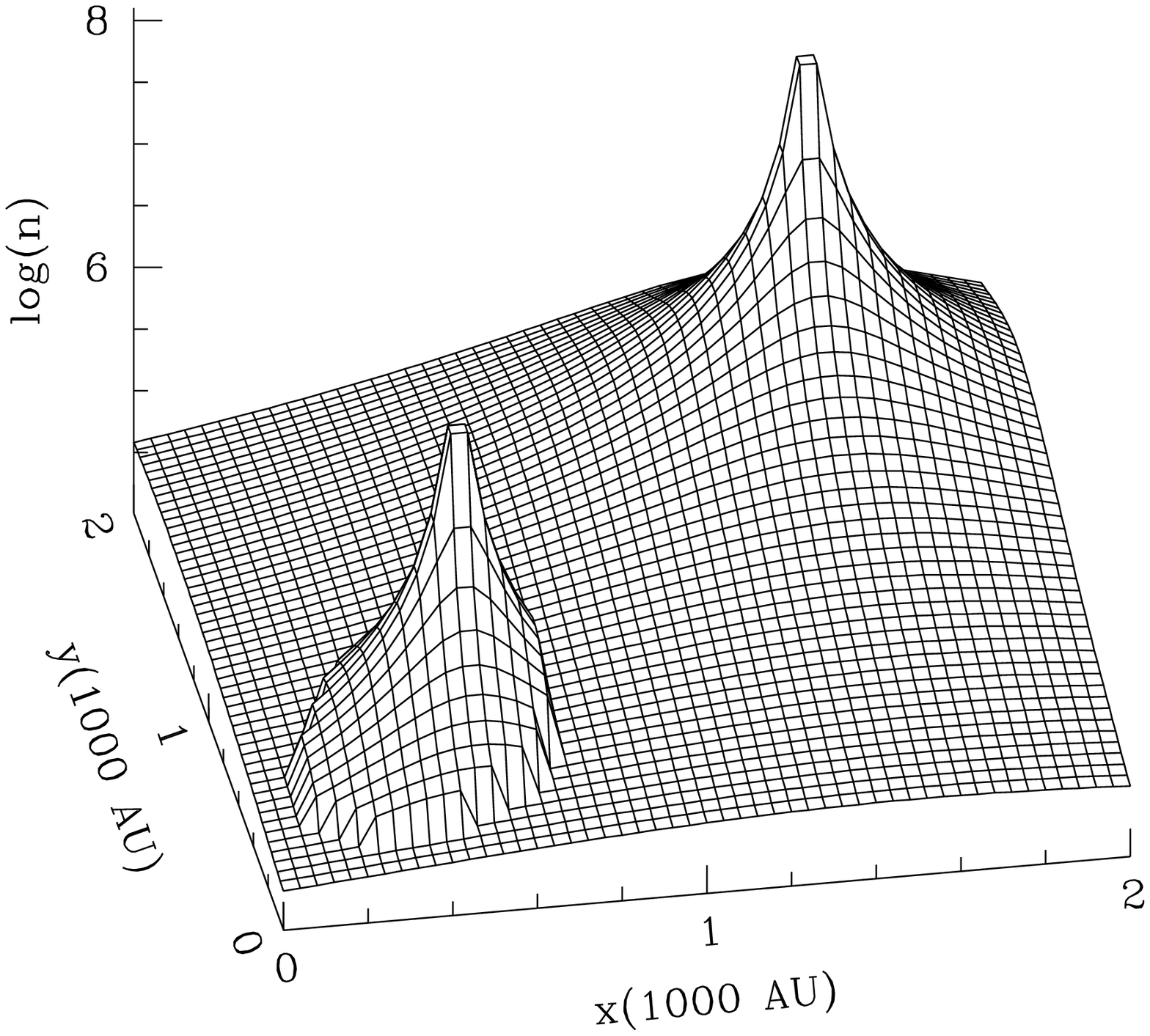}
\caption{Percentage difference between gas and dust temperature and density.  The same 
data and method as seen in Figures \ref{fig:tdust3} - \ref{fig:td32000surf}.  
First figure shows percentage difference of dust and gas.  Second figure shows
density.
\label{fig:perdiff}}
\end{figure}

We have presented a method for calculating the dust and gas temperature
between stellar sources. 
The analytic method that we investigated for calculating the dust temperature 
was not accurate enough.
Instead, our chosen method of calculating the dust temperature uses 
a simple radiative transfer code which we use to create a look-up table.
Once we have derived the dust temperature, we are able to calculate the
gas temperature by balancing various energy processes.  
We include dust-gas collisional heating, molecular cooling, and 
cosmic-ray heating.  
When we have balanced the energies, we are able to derive the gas
temperature.  
Other methods which set the gas temperature and dust
temperature equal assume the gas and dust are opaque everywhere, which
is not always true.  In Figure \ref{fig:perdiff}, we show the percentage
difference between the gas and dust temperature, as well as the density, 
for the distribution of sources discussed in \S \ref{sec:three}.  These two
figures show that the largest percentage difference between the dust
and gas temperature occurs where the density is the lowest.  Therefore, at 
low densities ($n \la 10^{5}$cm$^{-3}$), assuming equal 
dust and gas temperatures is not appropriate.

We plan to use the method discussed in this paper to model a region 
of clustered star formation with the three-dimensional hydrodynamics code 
discussed in \cite{martel}.
Our method of calculating the gas and dust temperature distribution 
in a field of young stars will enable us and others to more accurately model 
clustered star formation observationally and in future simulations.

\section{Acknowledgments}
AU would like to thank the NASA GSRP for providing support and 
Chad Young and Jeong-Eun Lee for help with DUSTY and the gas energetics code.
NE would like to thank the NSF for grants AST-0307250 and AST-0607793.
This work was partially supported by a grant from The Research Corporation (SDD).

\newpage 

\begin{deluxetable}{c | c | c c c }
\tabletypesize{\scriptsize}
\tablecaption{Dust Parameters}
\tablewidth{0pt}
\tablehead{
 \colhead{Dust Type}  & \colhead{$\frac{Q_a(UV)}{Q_e(125\mu m)}$} 
& \colhead{$K(1)$} 
& \colhead{$K(1.8)$} 
& \colhead{$K(2)$}
}
\startdata
OH5 (our dust model) & 253 & 539.3   & 351.1 & 319.8\\
\citet{hilde} & 4000 & 936.6 & 565.1 & 506.6 \\
\citet{makinen} & 790 &  677.1  & 427.2 & 386.7
\enddata
\label{tab:a}
\end{deluxetable}

\begin{deluxetable}{c | c c c c}
\tabletypesize{\scriptsize}
\tablecaption{Dust Model Parameters}
\tablewidth{0pt}
\tablehead{
 \colhead{Parameter}  & \colhead{Lower Limit} & \colhead{Upper Limit}& \colhead{$\Delta$\tablenotemark{a}} & \colhead{ N\tablenotemark{b}}
}
\startdata
log (L/$\Lsun$)  & -2 & 6 & 0.5 & 17\\
log ($n_o/1$cm$^{-3}$) & 2.5 & 7.5 & 0.5 & 11\\  
$\alpha$  & 0 & 4 &0.5 & 9\\
log (r$_{out}$/1 pc)  & -1 & 0 & 0.5  & 3
\enddata
\tablenotetext{a}{Spacing of parameters}
\tablenotetext{b}{Number of parameters}
\label{tab:2par}
\end{deluxetable}

\begin{deluxetable}{c c c c| c c c c c c}
\tabletypesize{\scriptsize}
\tablecaption{Dust Model Interpolation Results}
\tablewidth{0pt}
\tablehead{
 \colhead{L($\Lsun$)}  & \colhead{n$_o$ (cm$^{-3}$)} & \colhead{$\alpha$}& \colhead{$r_{out}$ (pc)}  
& \colhead{K$^{model}$} & \colhead{K$^{poly-interp}$}& \colhead{K$^{\%-diff}$}
& \colhead{$\beta^{model}$} & \colhead{$\beta^{poly-interp}$} & \colhead{$\beta^{\%-diff}$}}
\startdata
1.8$\ee{-2}$&5.62$\ee{2}$ & 1.75 & 0.177 &   302.1  & 300.8 & 0.4& 1.82 & 1.82  & 0.0\\
1.0$\ee{0}$ &5.00$\ee{3}$  & 2.75 & 0.194 &  214.8  & 213.9 & 0.4& 1.81 & 1.81  & 0.0  \\
1.0$\ee{0}$ &5.00$\ee{3}$  & 1.75 & 0.194 &  295.6  & 295.1 & 0.2& 1.78 & 1.78  & 0.0  \\
1.0$\ee{0}$& 5.00$\ee{2}$  & 0.80  & 0.194 &  322.4  & 310.0 & 3.8& 1.68 & 1.78  & 6.0 \\
1.0$\ee{0}$ & 5.50$\ee{5}$  & 2.00 & 0.194 &  161.5  & 157.6 & 2.4& 1.82 & 1.81  & 0.5 \\
1.5$\ee{5}$ & 5.00$\ee{2}$ & 2.00 & 0.194 &   355.8  & 356.5 & 0.2& 1.37 & 1.36  & 0.7 \\
\enddata
\label{tab:interp}
\end{deluxetable}

\begin{deluxetable}{c | c c c c | c c | c c}
\tabletypesize{\scriptsize}
\tablecaption{Source Parameters}
\tablewidth{0pt}
\tablehead{
 \colhead{Source}  & \colhead{Luminosity ($\Lsun$)}
& \colhead{$n_o$(cm${^{-3}}$}) 
& \colhead{$\alpha$} 
& \colhead{r$_{out}$ (pc)}
&{K} & {$\beta$} & x (AU) & y (AU)
}
\startdata
1  &     1   &  $10^3$ &   2   &    0.1  &    302.60 &  1.7875 & 500 & 1500\\
2  &     100 &  $10^5$ &   2   &    0.1  &    215.89  & 1.7623& 1500 & 1500\\
3  &     10  &  $10^4$ &   2   &    0.1  &    263.22 &  1.7817 & 500 & 500\\
\enddata
\label{tab:source}
\end{deluxetable}

\clearpage

\newpage 

\newpage 


\newpage
\begin{thebibliography}{}

 
\bibitem[Bate, Bonnell, \& Bromm(2003)]{bbb} Bate, M.~R., Bonnell, 
I.~A., \& Bromm, V.\ 2003, \mnras, 339, 577 

\bibitem[Churchwell(2002)]{Churchwell} Churchwell, E.\ 2002, 
\araa, 40, 27 

\bibitem[Doty \& Neufeld(1997)]{doty} Doty, S.~D., \& 
Neufeld, D.~A.\ 1997, \apj, 489, 122 

\bibitem[Flower \& Launay(1985)]{fl} Flower, D.~R., \& 
Launay, J.~M.\ 1985, \mnras, 214, 271 

\bibitem[Goldsmith(2001)]{goldsmith} Goldsmith, P.~F.\ 2001, 
\apj, 557, 736 

\bibitem[Hildebrand(1983)]{hilde} Hildebrand, R.~H.\ 1983, 
\qjras, 24, 267 

\bibitem[Hollenbach \& McKee(1989)]{hm} Hollenbach, D., \& 
McKee, C.~F.\ 1989, \apj, 342, 306 

\bibitem[Klessen, Burkert, \& Bate(1998)]{klessen} Klessen, R.~S., 
Burkert, A., \& Bate, M.~R.\ 1998, \apjl, 501, L205 

\bibitem[Krumholz, Klein, \& McKee(2007)]{krumholz} Krumholz, M.~R., 
Klein, R.~I., \& McKee, C.~F.\ 2007, \apj, 656, 959 

\bibitem[Lada \& Lada(2003)]{lada} Lada, C.~J., \& Lada, 
E.~A.\ 2003, \araa, 41, 57

\bibitem[Lee, Bergin, \& Evans(2004)]{lee} Lee, J.-E., Bergin, E.~A., 
\& Evans, N.~J., II 2004, \apj, 617, 360 

\bibitem[Makinen et al.(1985)]{makinen} Makinen, P., Harvey, 
P.~M., Wilking, B.~A., \& Evans, N.~J., II.\ 1985, \apj, 299, 341 

\bibitem[Martel et al.(2006)]{martel} 
Martel, H., Evans, N.~J., II, \& Shapiro, P.~R.\ 2006, \apjs, 163, 122 

\bibitem[McCaughrean \& O'dell(1996)]{mcC} McCaughrean, 
M.~J., \& O'dell, C.~R.\ 1996, \aj, 111, 1977 

\bibitem[Mueller et al.(2002)]{mueller} Mueller, K.~E., 
Shirley, Y.~L., Evans, N.~J., II, \& Jacobson, H.~R.\ 2002, \apjs, 143, 469 

\bibitem[Nenkova et al.(2000)]{dusty} Nenkova, M., 
Ivezi{\'c}, {\v Z}., \& Elitzur, M.\ 2000, ASP Conf.~Ser.~196: Thermal 
Emission Spectroscopy and Analysis of Dust, Disks, and Regoliths, 196, 77 

\bibitem[Neufeld \& Kaufman(1993)]{nk93} Neufeld, D.~A., \& 
Kaufman, M.~J.\ 1993, \apj, 418, 263 

\bibitem[Neufeld, Lepp, \& Melnick(1995)]{nlm95} Neufeld, D.~A., Lepp, 
S., \& Melnick, G.~J.\ 1995, \apjs, 100, 132 

\bibitem[Ossenkopf \& Henning(1994)]{oh5} Ossenkopf, V., \& 
Henning, T.\ 1994, \aap, 291, 943 

\bibitem[Press et al.(1992)]{numrec} Press, W.~H., Teukolsky, 
S.~A., Vetterling, W.~T., \& Flannery, B.~P.\ 1992, Cambridge: University 
Press, |c1992, 2nd ed.,  

\bibitem[Pollack et al.(1994)]{pollack} Pollack, J.~B., 
Hollenbach, D., Beckwith, S., Simonelli, D.~P., Roush, T., \& Fong, W.\ 
1994, \apj, 421, 615

\bibitem[Spitzer(1998)]{spitzer} Spitzer, L.\ 1998, Physical 
Processes in the Interstellar Medium, by Lyman Spitzer, Wiley-VCH, 
May 1998.

\bibitem[van der Tak \& van Dishoeck(2000)]{vans} van der 
Tak, F.~F.~S., \& van Dishoeck, E.~F.\ 2000, \aap, 358, L79 
 
\bibitem[Young \& Evans(2005)]{chad} Young, C.~H., \& Evans, 
N.~J.\ 2005, \apj, 627, 293 

\bibitem[Young et al.(2004)]{young} Young, K.~E., Lee, J.-E., 
Evans, N.~J., Goldsmith, P.~F., \& Doty, S.~D.\ 2004, \apj, 614, 252 
 
\end{thebibliography}
\end{document}